\documentclass[fleqn,usenatbib]{mnras}

\usepackage{newtxtext,newtxmath}
\usepackage{float}

\usepackage[T1]{fontenc}

\DeclareRobustCommand{\VAN}[3]{#2}
\let\VANthebibliography\thebibliography
\def\thebibliography{\DeclareRobustCommand{\VAN}[3]{##3}\VANthebibliography}

\usepackage{graphicx}	

\title[The True Core Mass Function of Protoclusters]{
Will ALMA Reveal the True Core Mass Function of Protoclusters? 
}

\author[P. Padoan et al.]{
P. Padoan,$^{1,2}$\thanks{E-mail: ppadoan@icc.ub.edu} 
V.-M. Pelkonen,$^{1}$
M. Juvela,$^{3}$
T. Haugb{\o}lle$^{4}$
and \AA. Nordlund$^{4}$
\\
$^{1}$Institut de Ci\`{e}ncies del Cosmos, Universitat de Barcelona, IEEC-UB, Mart\'{i} i Franqu\'{e}s 1, E08028 Barcelona, Spain\\
$^{2}$ICREA, Pg. Llu\'{i}s Companys 23, 08010 Barcelona, Spain\\
$^{3}$ Helsinki......
$^{4}$Niels Bohr Institute, University of Copenhagen, {\O}ster Voldgade 5-7, DK-1350 Copenhagen, Denmark
}

\date{Accepted XXX. Received YYY; in original form ZZZ}

\pubyear{2022}

\begin{document}
\label{firstpage}
\pagerange{\pageref{firstpage}--\pageref{lastpage}}
\maketitle

\begin{abstract}
Characterizing prestellar cores in star-forming regions is an important step towards the validation of theoretical models of star formation. Thanks to their sub-arcsecond resolution, ALMA observations can potentially provide samples of prestellar cores up to distances of a few kpc, where regions of massive star formation can be targeted. However, the extraction of real cores from dust-continuum observations of turbulent star-forming clouds is affected by complex projection effects. In this work, we study the problem of core extraction both in the idealized case of column-density maps and in the more realistic case of synthetic 1.3\,mm ALMA observations. The analysis is carried out on 12 regions of high column density from our 250 pc simulation. We find that derived core masses are highly unreliable, with only {\em a weak correlation between the masses of cores selected in the synthetic ALMA maps and those of the corresponding three-dimensional cores}. The fraction of real three-dimensional cores detected in the synthetic maps increases monotonically with mass and remains always below 50\%. Above $\sim 1\,M_{\odot}$, the core mass function derived from the column-density maps is steeper than that of the three-dimensional cores, while the core mass function from the synthetic ALMA maps has a slope closer to that of the real three-dimensional cores. Because of the mass uncertainties, proper guidance from realistic simulations is essential if ALMA observations of protoclusters at kpc distances are to be used to test star-formation models.         
\end{abstract}



\section{Introduction}

The relation between the mass of a prestellar core and that of the star it forms is an important prediction of theoretical models of star formation. On the one hand, in the core-collapse model \citep[][]{McKee+Tan02,McKee+Tan03} and in some models of the initial mass function (IMF) of stars \citep[e.g.][]{Hennebelle+Chabrier08imf,Hopkins12imf}, it is assumed that the stellar mass reservoir is fully contained in a bound prestellar core. On the other hand, in the competitive accretion model \citep{Zinnecker82,Bonnell+2001a,Bonnell+2001b}, most of the stellar mass is accreted over time from a larger gas reservoir shared with other stars, with an accretion rate that depends on the mass of the star. Numerical simulations of star formation under turbulent conditions consistent with Larson's velocity-size relation \citep[e.g.][]{Larson81,Heyer+Brunt04} and with virial parameter of order unity seem to rule out both models \citep{Padoan+14acc,Pelkonen+21} and to suggest an alternative scenario, the {\em inertial-inflow model}, where the prestellar core contains only a fraction of the final stellar mass, while the remaining mass is brought to the star from larger scale by preexisting converging flows \citep{Padoan+20massive,Pelkonen+21}. The IMF model of \citet[][]{Padoan+Nordlund02imf} is consistent with this alternative scenario, as the final stellar mass is determined by preexisting converging flows, and can be much larger than the critical mass, particularly towards larger stellar masses \citep{Padoan+Nordlund11imf}.  

Observations do not directly constrain the relation between the final mass of a star and the mass of its prestellar core; they only offer us a single time snapshot with an ensemble of cores and young stars. The masses of cores and stars can only be related statistically, by comparing the core mass function (CMF) with the stellar IMF \citep{Salpeter55,Kroupa01,Chabrier05}. Observations of nearby star-forming regions seem to indicate that the CMF has a similar shape as the stellar IMF, which is often interpreted as evidence in favor of the core-collapse model \citep[e.g.][]{Motte+98,Alves+07,Enoch+07,Nutter+Ward-Thompson07,Konyves+10,Konyves+15,Marsh+16,Sokol+19,Konyves+2020_orion,Ladjelate+2020_ophiucus,Takemura+21,Takemura+22}. However, in more distant regions of high-mass star formation, {\em Herschel}'s observations \citep{Tige+17} and, more conclusively, interferometric surveys \citep[e.g.][]{Sanhueza+17,Sanhueza+19,Li+19,Pillai+19,Kong19cmf,Servajean+19} have revealed a scarcity of massive prestellar cores, implying that the mass reservoir to form high-mass stars is spread over larger scales, in contrast with the core-collapse model. To further complicate the picture, in some of the most extreme regions of high-mass star formation, the CMF at intermediate and large masses is sometimes found to be a shallower power law than the stellar IMF \citep{Motte+18,Pouteau+22}, which would imply either core fragmentation or a shallower IMF in such regions.   
The observational constraints on theoretical models rely on the statistical significance of the estimated values of the turnover mass and slope of the CMF. These values can be quite uncertain, as the turnover mass is often close to the completeness limit of the surveys and both turnover and slope can depend on the core-selection algorithm and may be affected by projection artifacts. {\em The goal of this work is to assess the reliability of CMFs derived from ALMA dust continuum observations of Galactic protoclusters}. As a test bed, we use our 250\,pc star-formation simulation driven self-consistently by supernova (SN) explosions \citep{Padoan+16SN_I,Pan+Padoan+SN2+2016ApJ,Padoan+SN3+2016ApJ,Padoan+17_SN4}. The maximum resolution of the simulation is 0.0076 pc, adequate to address ALMA observations of Galactic protoclusters at a resolution of order 0.01\,pc, such as those of the ALMA-IMF Large Program \citep{Motte+22alma_imf,Ginsburg+22,Pouteau+22}. Although the turnover mass of the CMF is not fully resolved at this resolution \citep[e.g.][]{Pelkonen+21}, the simulation is realistic enough to allow us to test the relation between cores selected from two-dimensional (2D) projections, as in the observations, and real cores selected from the corresponding three-dimensional (3D) volumes, irrespective of the actual value of the turnover mass.  

We first analyze column-density maps obtained from the simulation, and then synthetic 1.3 mm dust-continuum maps computed with the radiative transfer code SOC \citep{Juvela_2019}. We aim at reproducing a synthetic ALMA dataset and an analysis pipeline as close as possible to that of the ALMA-IMF Large Program \citep{Motte+22alma_imf,Ginsburg+22,Pouteau+22}. As in \citet{Pouteau+22}, the cores are extracted with the {\em getsf} code \citep{Menshchikov21}. Cores are also extracted from the corresponding 3D volumes with a dendrogram analysis \citep{Padoan+07imf,Rosolowsky+08} and compared with the sample of 2D cores from {\em getsf}. The comparison shows that the 2D cores present a highly incomplete view of the sample of real 3D cores, are sometimes observational artifacts from projection effects, and have masses poorly correlated with those of the corresponding 3D cores. 

The paper is organized as follows. In the next section we briefly summarize the simulation and in \S~\ref{column_density} we describe the selection and the general properties of the column density maps. The computation of the synthetic observations is presented in \S~\ref{synthetic_maps} and the extraction of cores in both 2D and 3D is described in \S~\ref{getsf_clumpfind}. The results from the analysis of both column-density maps and synthetic observations is presented in \S~\ref{results}. A discussion of our results, and their significance with respect to recent findings from the ALMA-IMF Large Program, is given in \S~\ref{discussion}, and the main conclusions are summarized in \S~\ref{conclusions}.

  \begin{figure*}
   \centering
   \includegraphics[width=4.37cm]{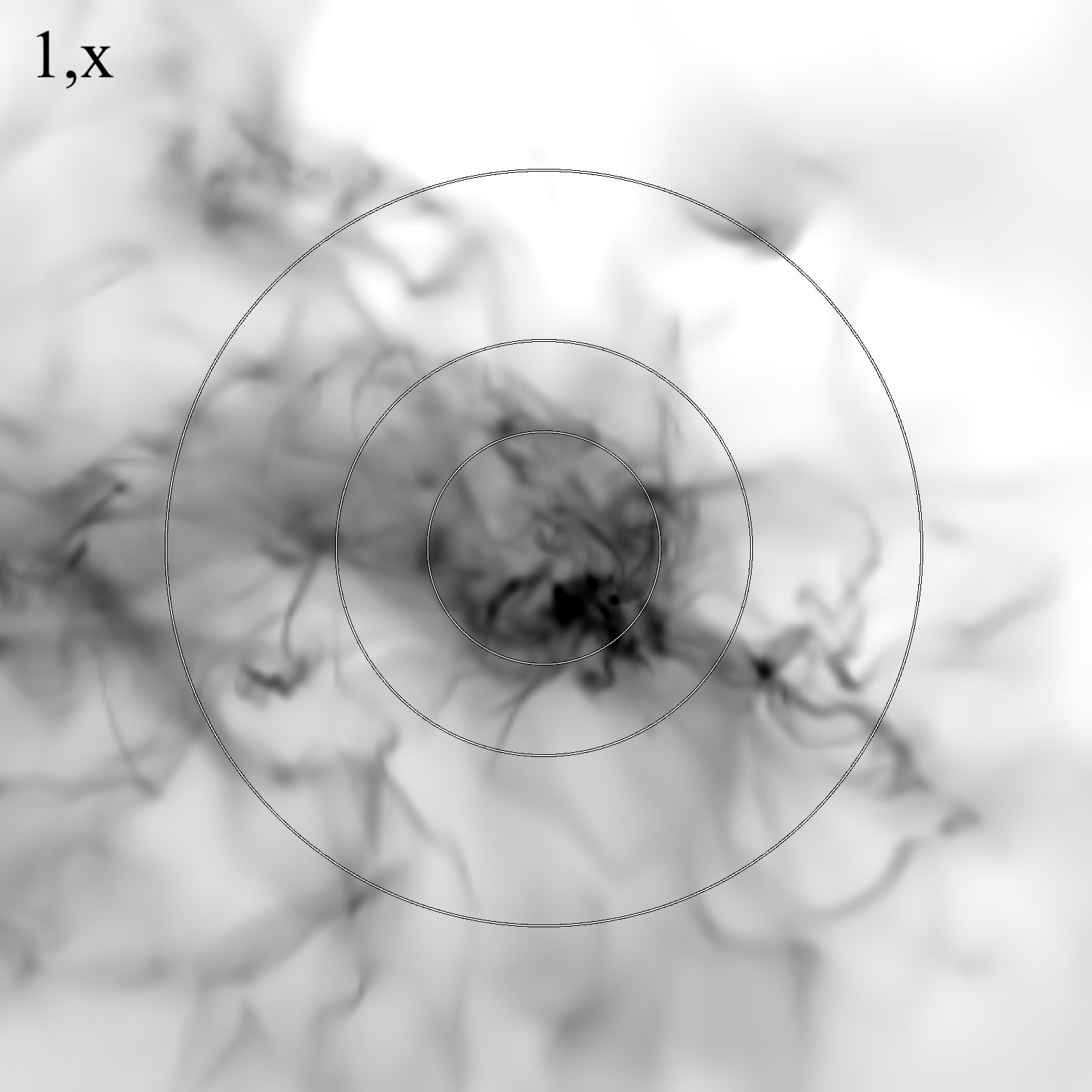}
   \includegraphics[width=4.37cm]{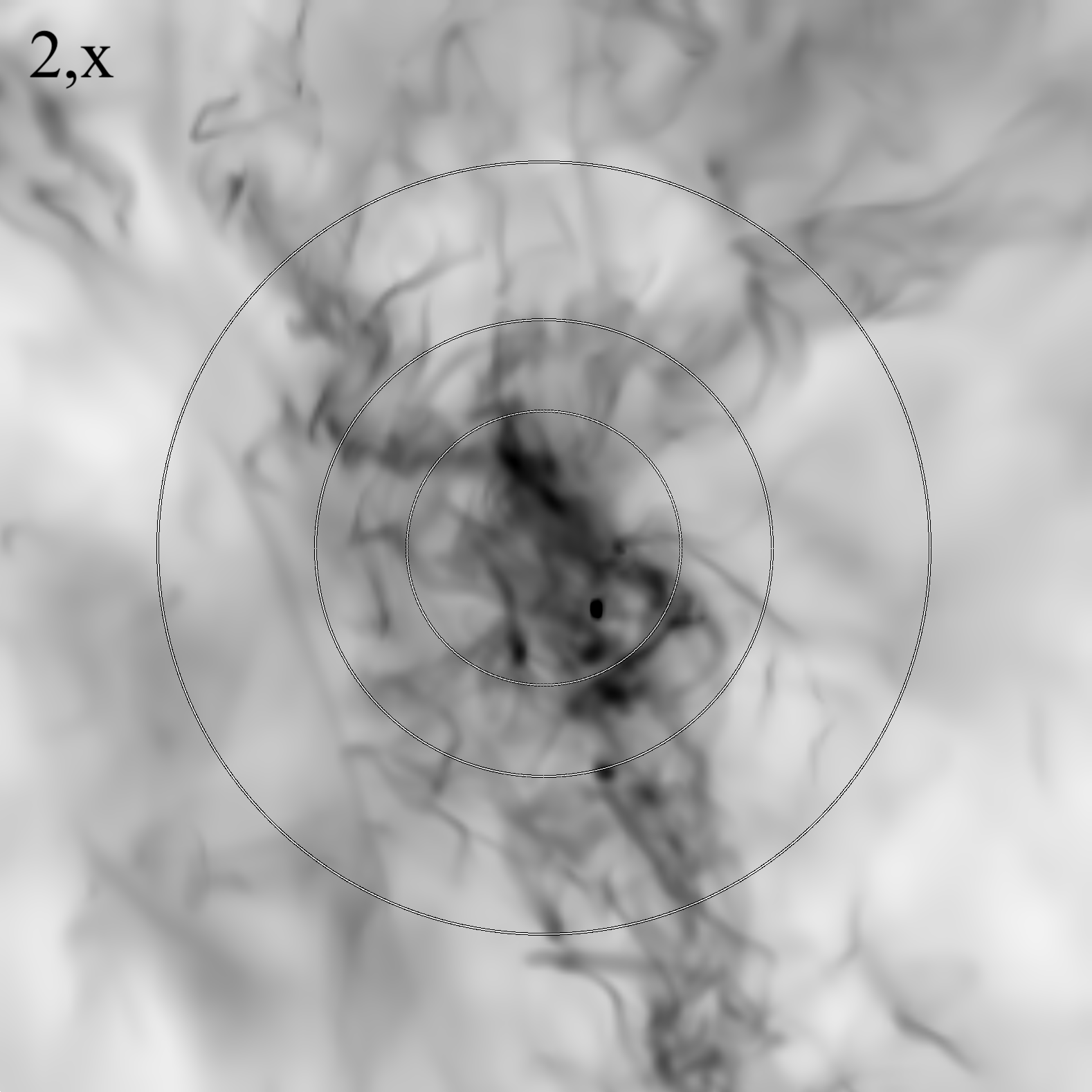}
   \includegraphics[width=4.37cm]{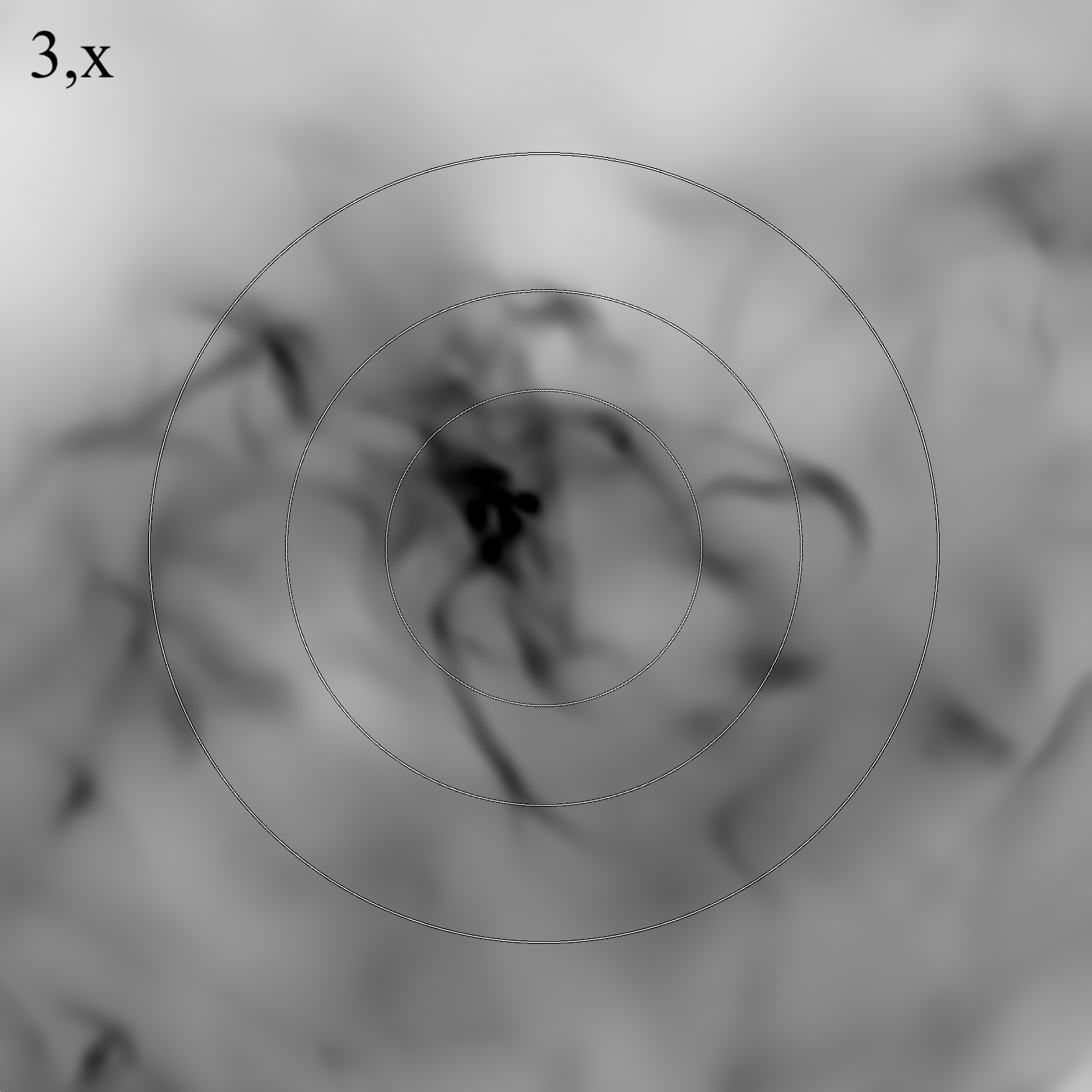}
   \includegraphics[width=4.37cm]{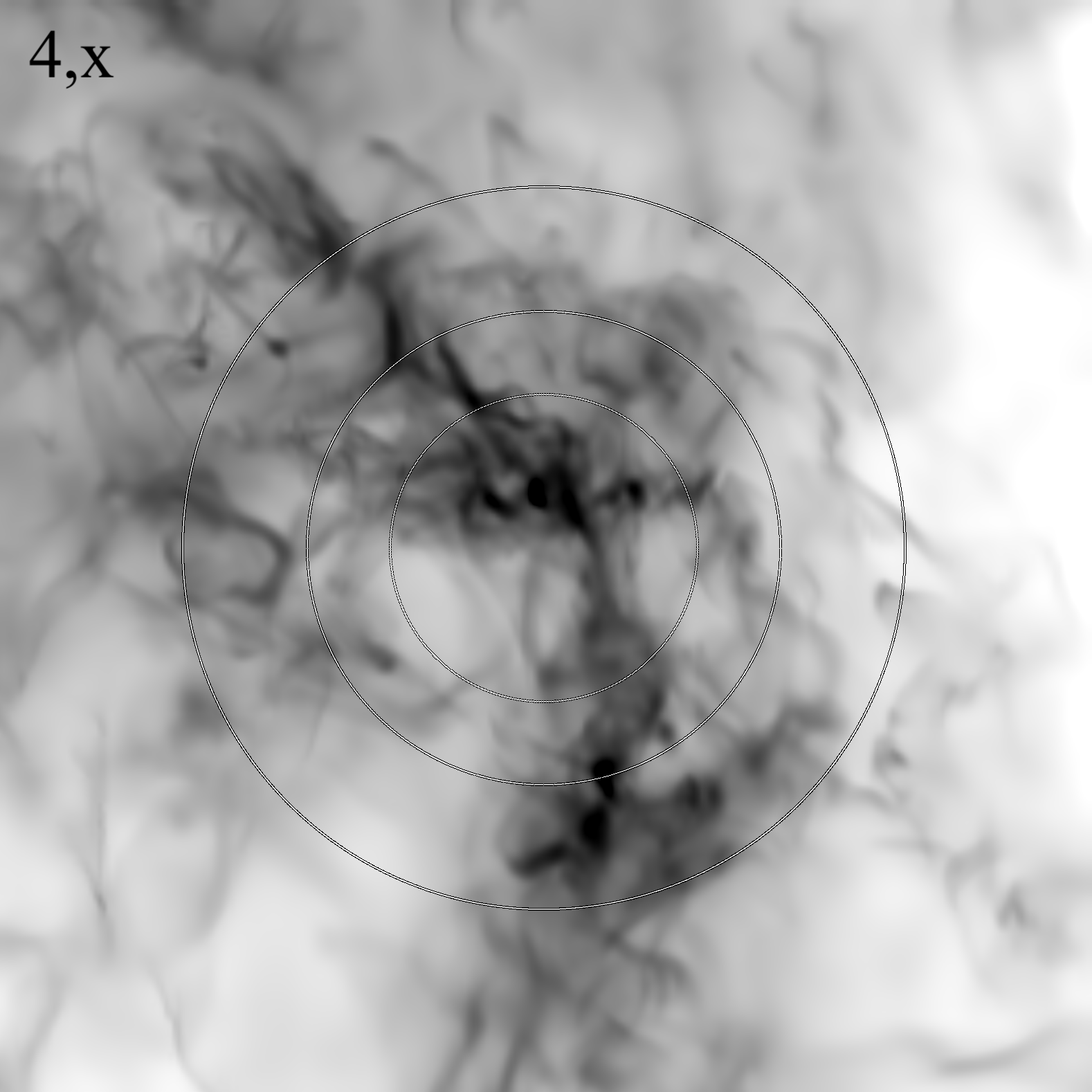}
   \includegraphics[width=4.37cm]{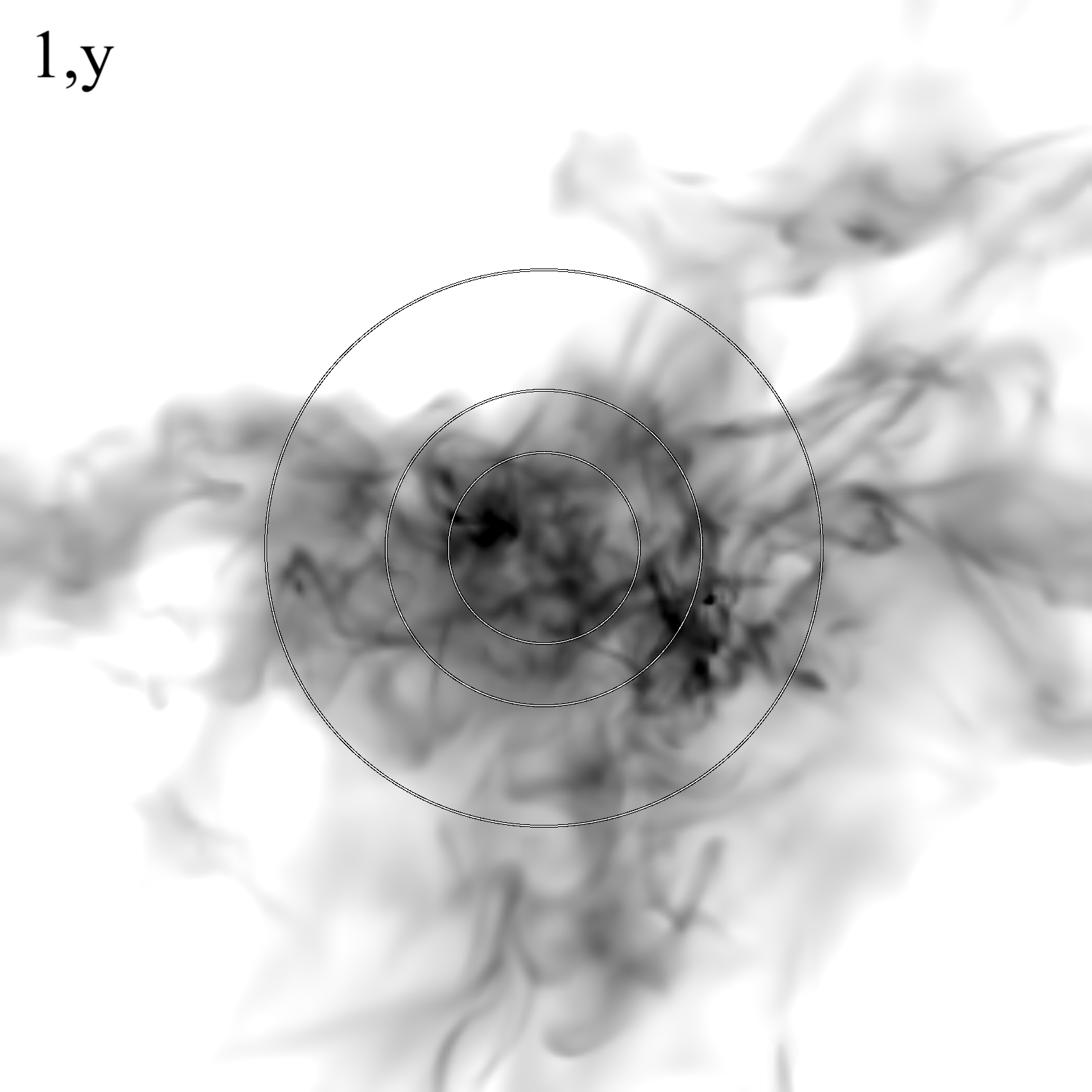}
   \includegraphics[width=4.37cm]{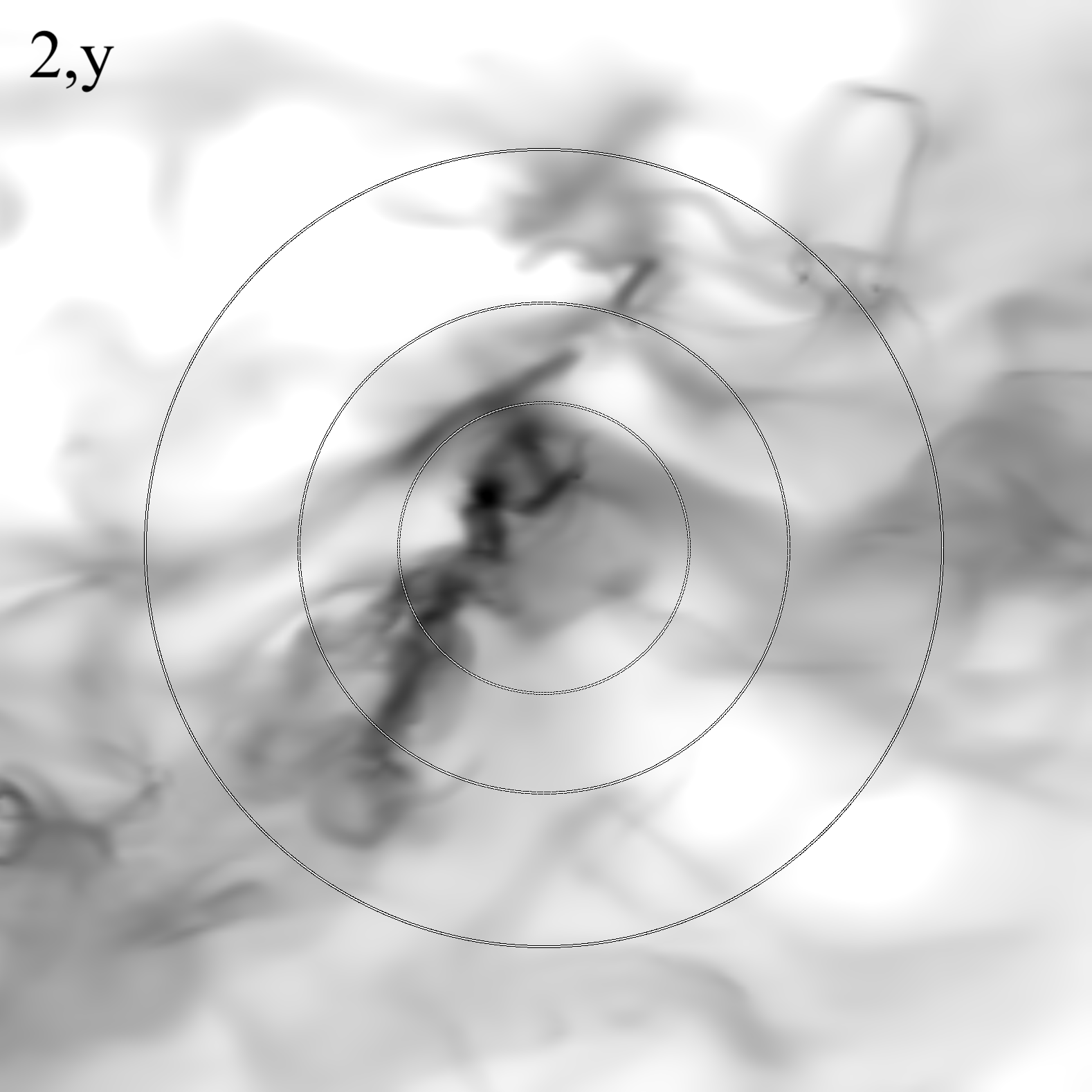}
   \includegraphics[width=4.37cm]{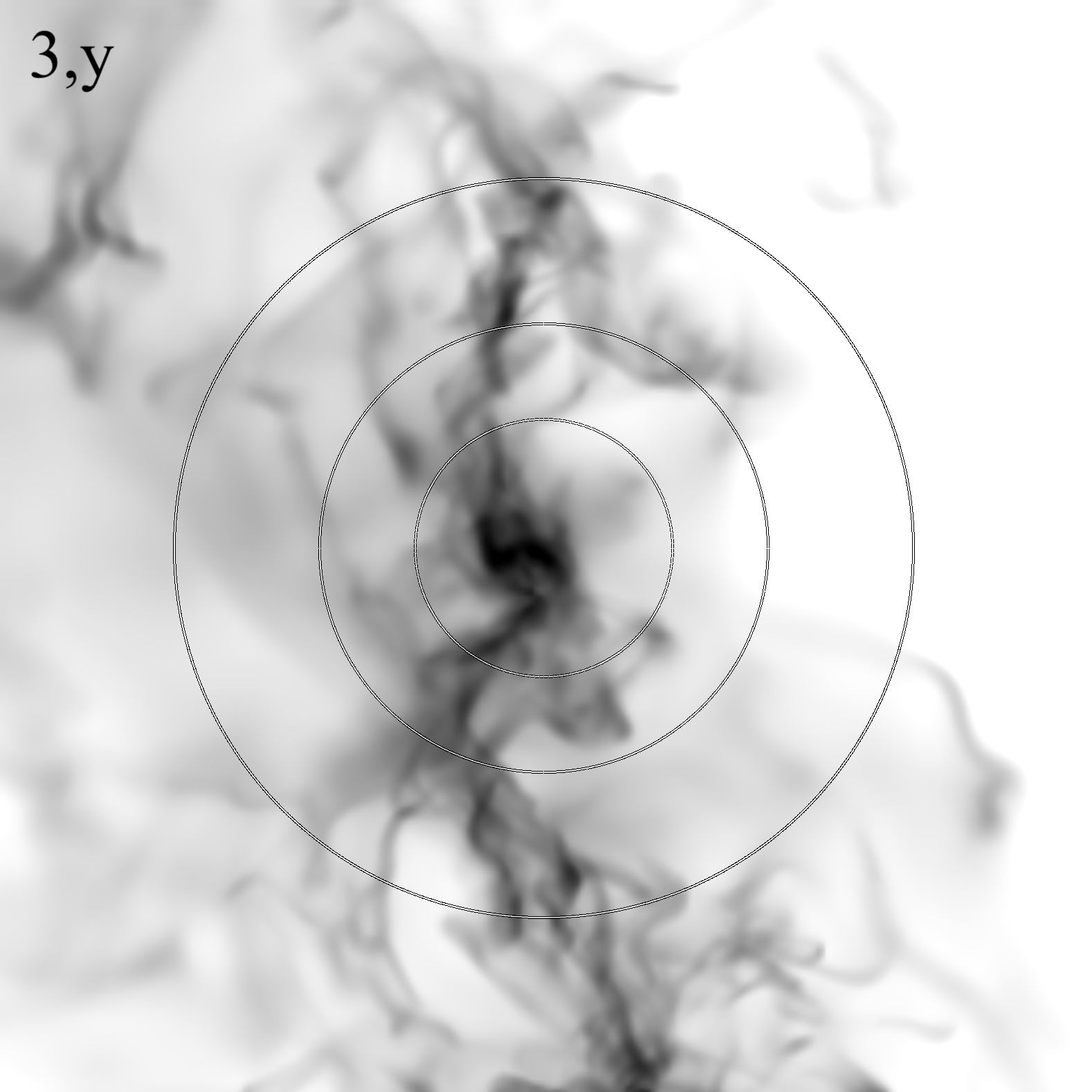}
   \includegraphics[width=4.37cm]{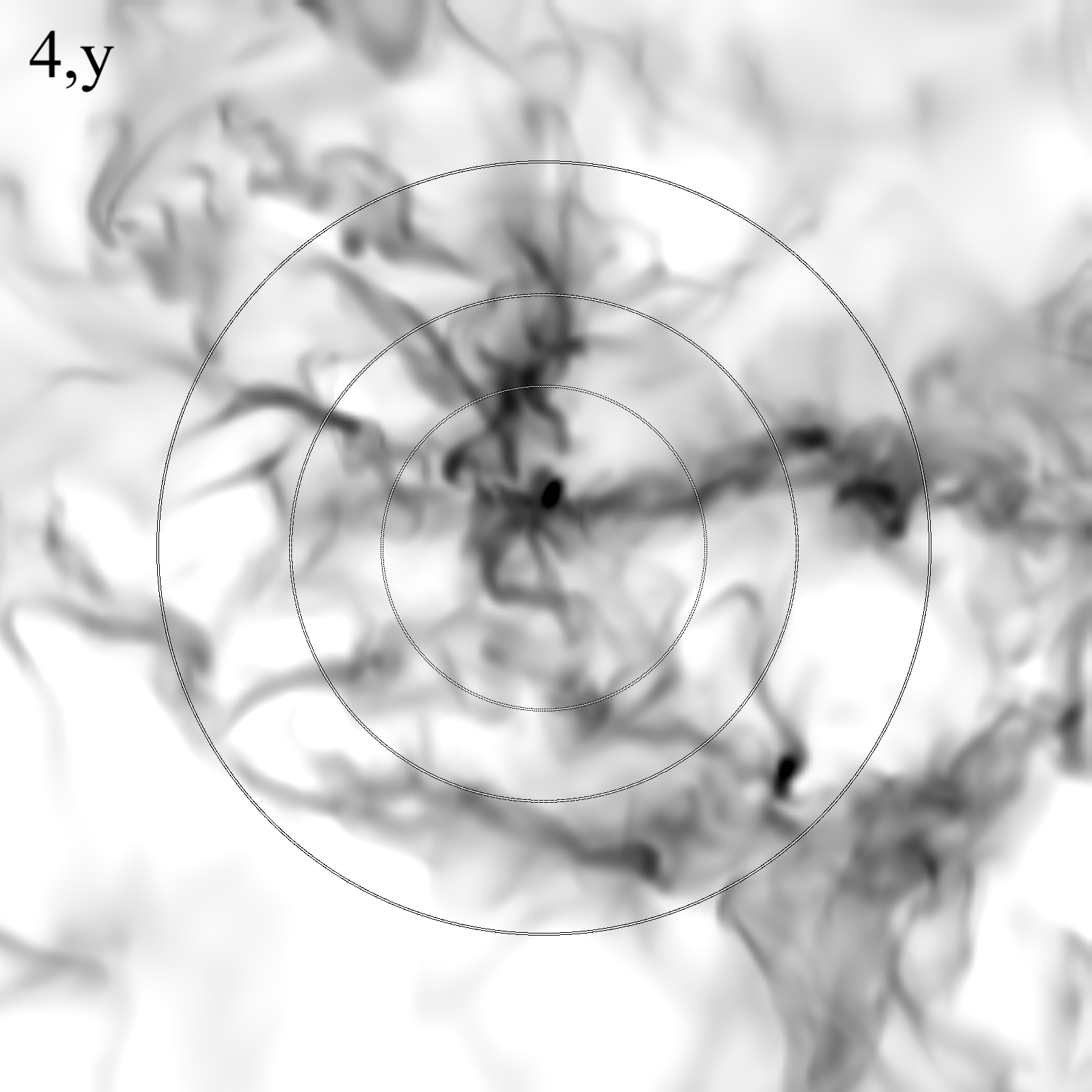}
   \includegraphics[width=4.37cm]{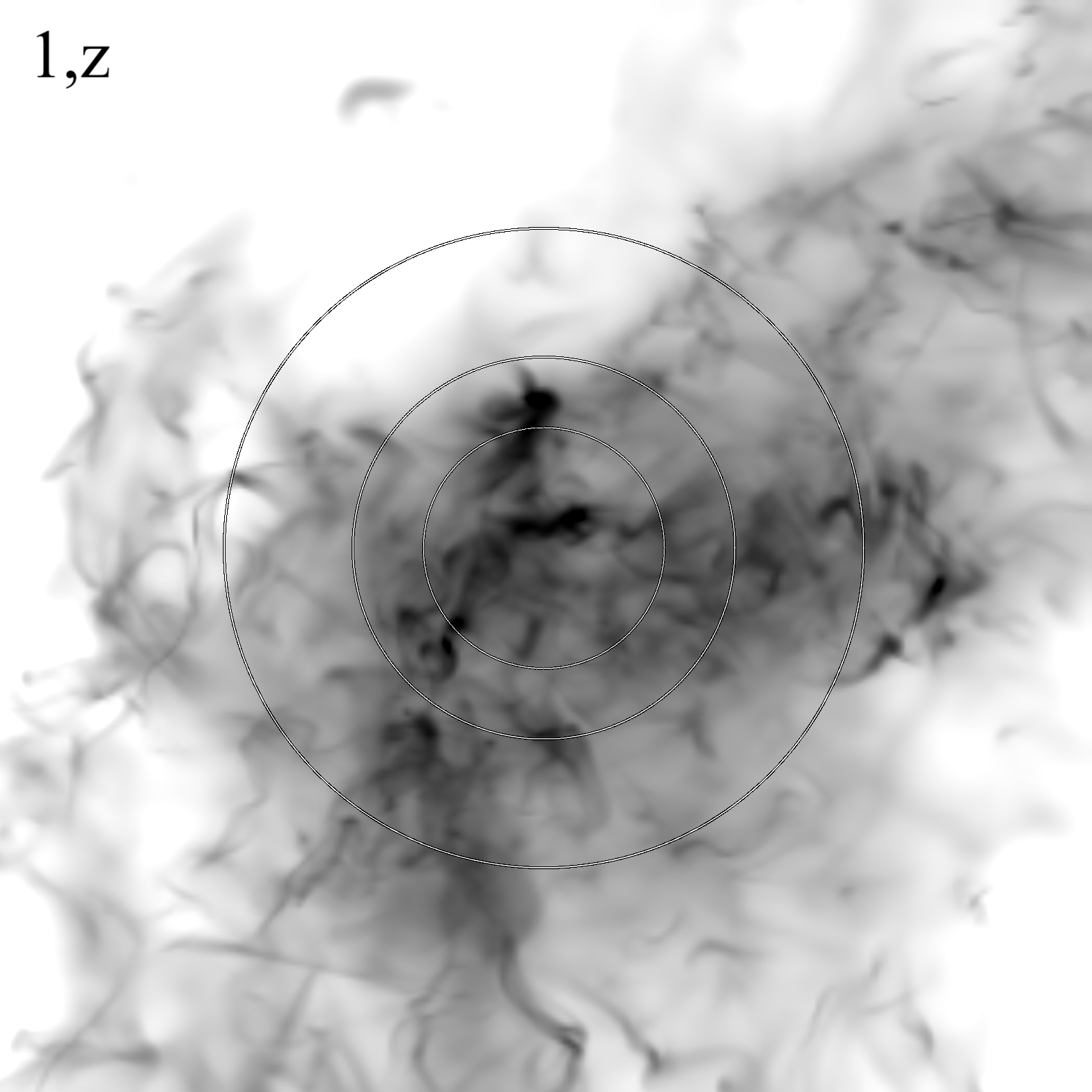}
   \includegraphics[width=4.37cm]{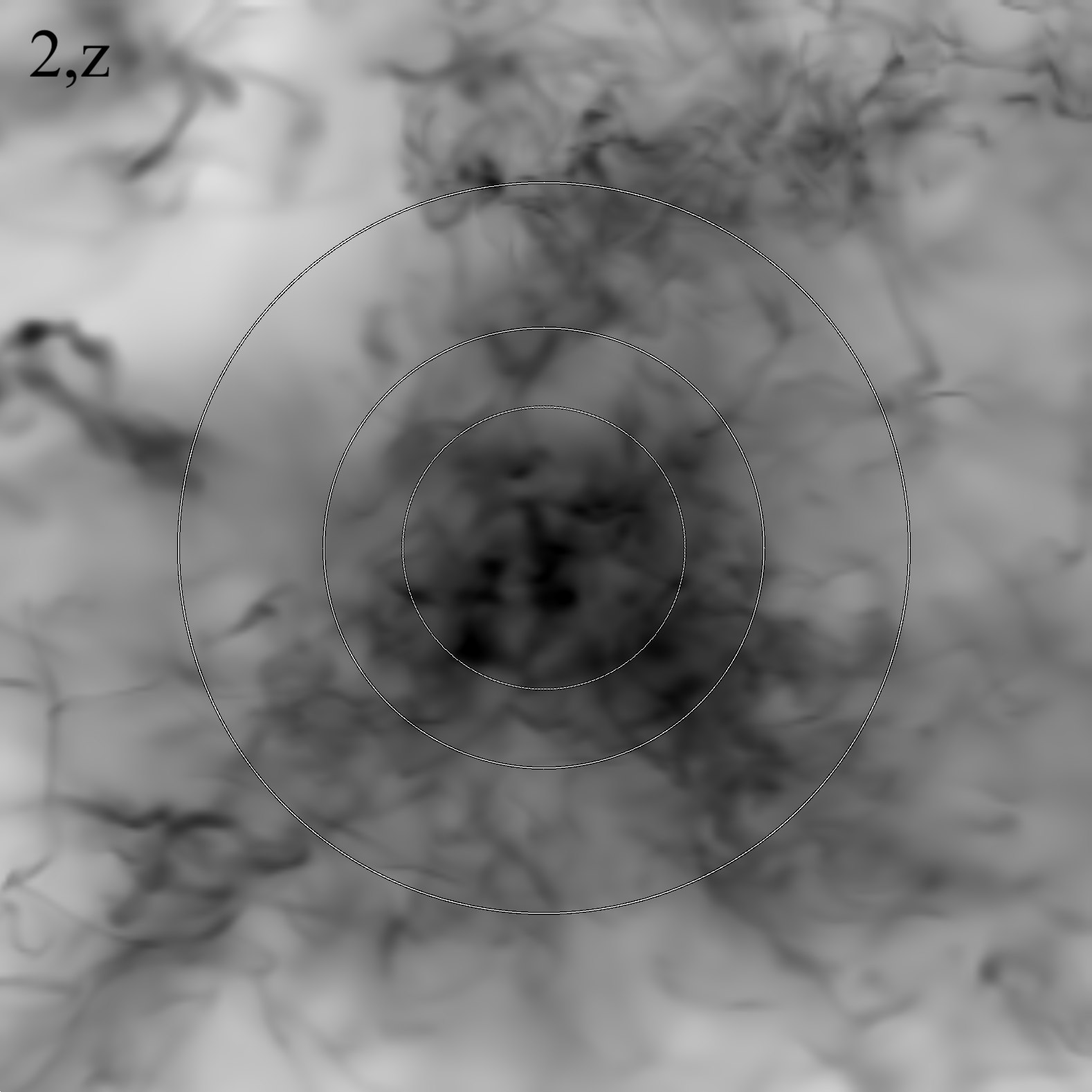}
   \includegraphics[width=4.37cm]{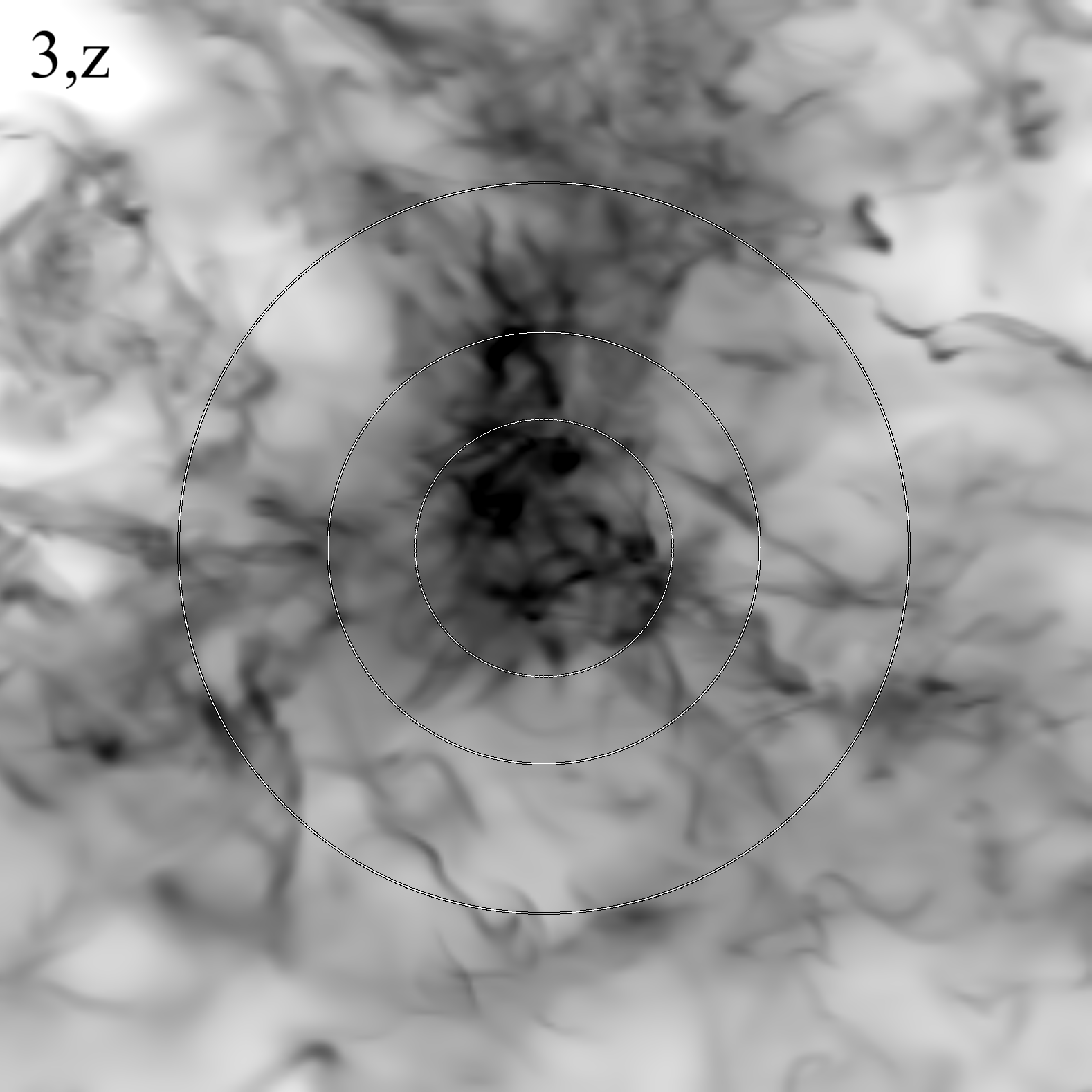}
   \includegraphics[width=4.37cm]{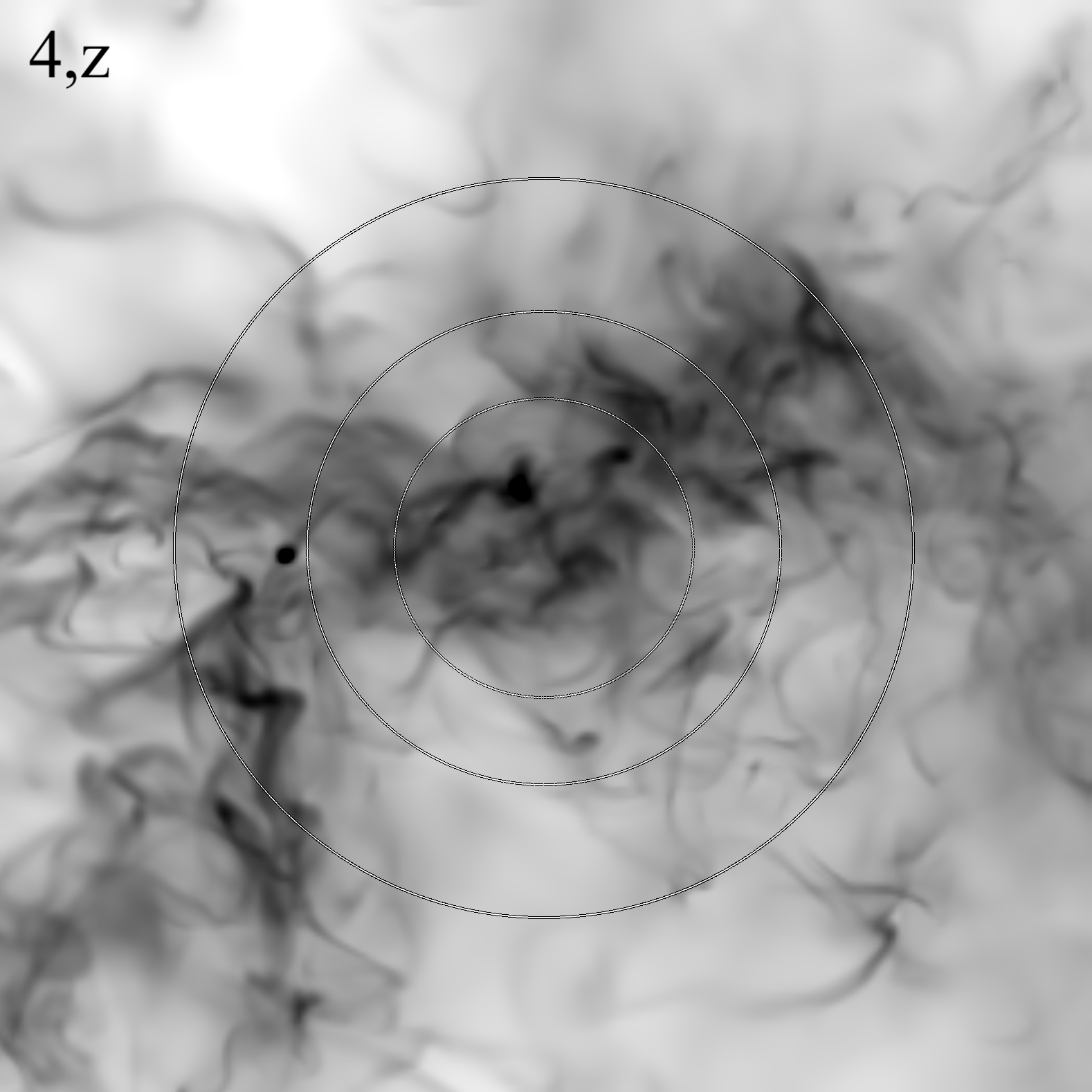}
      \caption{Column-density maps of the 12 regions analyzed in this work. Each map covers a 2~pc~$\times$~2~pc region centered around the highest column density of the corresponding projection of the whole 250~pc computational volume (see text for details). The four columns of panels correspond to the four different snapshots of the simulation taken at time intervals of nearly 7~Myr; the three rows of panels correspond to the three orthogonal directions. The circles show the areas containing 12.5\%, 25\% and 50\% of the total mass of each map. The grey scale is proportional to the square root of the column density, with minimum and maximum values set to 0.05~g~cm$^{-2}$ (white colour) and 1.0~g~cm$^{-2}$ (black colour) respectively.
              }
         \label{images_small}
   \end{figure*}

  \begin{figure}
   \centering
   \includegraphics[width=\hsize]{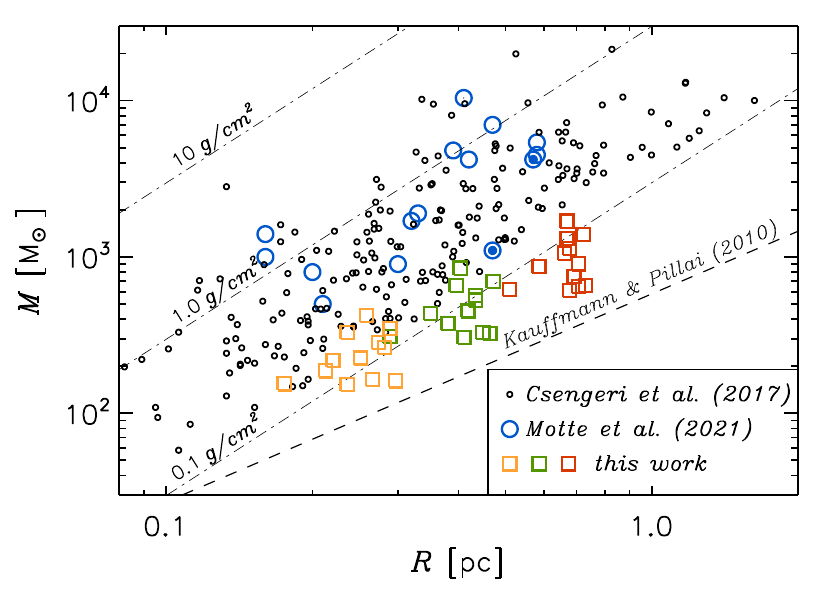}
         \caption{Mass versus size of the 12 regions analyzed in this work (squared symbols). The three clusters of symbols show the values of radius and mass in the case where 12.5\% (yellow squares), 25\% (green squares), and 50\% (red squares) of the total mass of each map is considered (see the three circles in the panels of Figure~\ref{images_small}). The small, black circles correspond to the most luminous ATLASGAL regions from Csengeri et al. (2017), and the larger blue circles the masses and sizes of the 15 ALMA-IMF regions in \citet{Motte+22alma_imf}. The two open blue circles with a filled circle inside correspond to W43-MM2 (higher mass) and W43-MM3 (lower mass) studied in \citet{Pouteau+22}. The dashed line is the empirical mass-size relation limit for massive star formation of \citet{Kauffmann+2010}, as revised by \citet{Dunham+11}, and the dashed-dotted lines correspond to different values of constant column density.
              }
         \label{mass_size}
   \end{figure}

\section{Simulation} \label{simulation}

The test bed for this work is a sample of protocluster regions found in our 250\,pc star-formation simulation driven self-consistently by SNe. The simulation has been continuously run, during the past three years, under a multi-year PRACE\footnote{Partnership for Advanced Computing in Europe} project, until it reached approximately 45~Myr of evolution under self-gravity. It describes an ISM region of size $L_{\rm box}=250$ pc and total mass $M_{\rm box}=1.9\times 10^6$ $\rm M_{\odot}$, where the turbulence is driven by SNe alone. Given the large time and spatial scales, the simulation develops many regions where high-mass stars are formed, including several stellar clusters. 

The simulation has been shown to generate star-forming clouds with realistic observational properties, including kinematics, lifetimes, and star-formation rates \citep{Padoan+16SN_I,Padoan+16SN_III,Lu+20SN}. It has also been used to study the statistical properties of SN-driven turbulence \citep{Padoan+16SN_I,Pan+Padoan+SN2+2016ApJ}, to propose the new {\em Inertial-Inflow} scenario for the formation of massive stars \citep{Padoan+17_SN4}, and to assess the real nature of massive clumps from the {\em Hi-GAL} Survey \citep{Lu+2022_higal}. The reader is referred to \citet{Padoan+16SN_I,Padoan+16SN_III} for details about the numerical setup. The main features of the simulation relevant to this work are briefly summarized in the following. 

The 3D MHD equations are solved with the adaptive-mesh-refinement (AMR) code RAMSES \citep{Teyssier+2002A&A,Fromang+06,Teyssier07}, using periodic boundary conditions. The energy equation includes the pressure-volume work, the thermal energy introduced to model SN explosions, a uniform photoelectric heating as in \citet{Wolfire+95}, with efficiency $\epsilon=0.05$ and the FUV radiation field of \citet{Habing68} with coefficient $G_0=0.6$ (the UV shielding in MCs is approximated by tapering off the photoelectric heating exponentially above a number density of 200 cm$^{-3}$), and a tabulated optically thin cooling function constructed from the compilation by \citet{Gnedin+Hollon12} that includes all relevant atomic transitions. Molecular cooling is not included, due to the computational cost of solving the radiative transfer. The thermal balance between molecular cooling and cosmic-ray heating in dense gas is emulated by setting a limit of 10~K as the lowest temperature of dense gas. However, to generate synthetic observations of the dust emission, the radiative transfer is computed by postprocessing individual snapshots, including all stars with mass $>2 \, \rm M_{\odot}$ as point sources (see \S~\ref{synthetic_maps}).     

The simulation is initialized with zero velocity, uniform density, $n_{\rm H,0}=5$\,cm$^{-3}$, uniform temperature, $T_0=10^4$\,K, and uniform magnetic field, $B_0=4.6$ $\mu$G. During the first 45\,Myr, self-gravity was not included and SN explosions were randomly distributed in space and time, at a rate of 6.25 SNe Myr$^{-1}$. The resolution was $dx=0.24$ pc, achieved with a $128^3$ root grid and three AMR levels. These conditions are chosen to represent a generic Galactic spiral arm region, and the timescale of 45\,Myr is long enough to achieve a statistical steady state, prior to the inclusion of gravity. The first few SN explosions rapidly bring the mean thermal, magnetic and kinetic energies to approximately steady-state values, with an rms magnetic field of 7.2\,$\mu$G, and an average of $\vert \boldsymbol{B} \vert$ of 6.0\,$\mu$G, consistent with the value of $6.0 \pm 1.8$\,$\mu$G derived from the `Millennium Arecibo 21-cm Absorption-Line Survey' by \citet{Heiles+Troland05}. The minimum cell size was then decreased to $dx=0.03$\,pc, using a root-grid of $512^3$ cells and four AMR levels, for an additional period of 10.5 Myr, still without self-gravity. At $t=55.5$\,Myr, gravity is introduced and the simulation is continued until $t\approx 100$\,Myr with a minimum cell size further reduced to $dx=0.0076$ pc, by adding two more AMR levels. At this resolution, we can follow the formation of individual massive stars, so the time and location of the SNe are computed self-consistently from the evolution of those stars. 

Individual stars are modeled with accreting sink particles, created when the gas density is larger than $10^6$ cm$^{-3}$ and other conditions are satisfied \citep[see][for details of the sink particle model]{Haugboelle+2018}. A SN is created when a sink particle of mass larger than 7.5~$\rm M_{\odot}$ has an age equal to the corresponding stellar lifetime for that mass \citep{Schaller+92}. The sink particle is removed and the stellar mass, momentum, and $10^{51}$ erg of thermal energy are added to the grid with a Gaussian profile \citep[see][for further details]{Padoan+SN1+2016ApJ}. By the latest simulation snapshot used in this work, corresponding to a time of 34.7\,Myr from the inclusion of self-gravity and star formation, 3,942 stars with mass $> 2\,M_{\odot}$ have been generated, of which 389 have already exploded as SNe. The stellar mass distribution is consistent with Salpeter's IMF \citep{Salpeter55} above $\sim 8\,\rm M_{\odot}$, but is incomplete at lower masses (it starts to flatten at a few solar masses instead of at a fraction of a solar mass), as expected for the spatial resolution of the simulation.

\section{Column Density Maps} \label{column_density}

The goal of this work is to study simulated star-forming regions that are comparable to observed regions of high-mass star formation. For that purpose, we select the highest column density regions from four different snapshots of the simulation, spanning a range of times between 14.2 and 34.7\,Myr from the beginning of self-gravity. We generate maps of 2\,pc\,$\times$\,2\,pc, as that is a characteristic size of the 15 regions mapped by the ALMA-IMF Large Program. For each of the four snapshots, we find the position of maximum column density measured at a resolution of 512$^2$ cells, or 0.49\,pc \citep[a characteristic size of massive clumps in][]{Csengeri+2017} in the three orthogonal directions of the 250\,pc computational volume. That results in 12 positions of maximum column density, which are taken as the centers of 12 maps of 2\,pc\,$\times$\,2\,pc. The column density maps are then recomputed at the maximum resolution of the simulation, 0.0076\,pc, so each map is composed of 263\,$\times$\,263 pixels. 

We stress that these maps result from the projection of 3D columns with a very large aspect ratio of 125:1 (250\,pc\,$\times$\,2\,pc\,$\times$\,2\,pc); while they are meant to represent relatively small star-forming regions, as those modelled with smaller simulations, they are embedded in a realistic 250\,pc volume, and they contain the full projection effects to be expected at that scale. As extensively discussed in \citet{Lu+2022_higal}, this projection depth is representative of the thickness and total column density of a characteristic spiral arm. Observations of similar regions in the Galactic plane and towards the Inner Galaxy may suffer much larger projection effects, as they may sample several dense regions along a distance of several kpc, rather than structures within a single spiral arm. This should be kept in mind when comparing our simulated regions with observations of the highest column density regions in the Galaxy, as discussed in the following.  

The 12 maps from the simulation are truly independent of each other, even if they are selected from only four snapshots. We have verified that only a small fraction of the 3D columns in each snapshot intersect each other. Of these, only one intersection, between directions $y$ and $z$ of snapshot 3, corresponds to a 3D volume whose mass gives a significant contribution to the column density in the maps. However, while the map of direction $z$ contains most of the mass of the map in direction $y$, the map of direction $y$ contains less than half of the mass of the map in direction $z$. Thus, even these two maps cannot be considered two different views of the same 3D region.  Figure~\ref{images_small} shows the 12 column density maps. The intersecting ones are shown by panels $3,y$ and $3,z$. The map in panel 3,y shows a dominant dense filament along the $z$ direction that seems to continue beyond the bottom of the map. The map in panel $3,z$ shows the projection of that filament in the $z$ direction, and so it also includes the part of the filament that lies beyond the bottom of the map in panel $3,y$.

\subsection{Simulated Maps as Protocluster Regions} \label{protoclusters}

All the maps in Figure~\ref{images_small} appear like clusters of filaments, with the highest column densities corresponding to regions where many filaments intersect. This morphology is very similar to that of the so-called filament hubs that are ubiquitous in regions of high-mass star formation \citep[e.g.][]{Myers09_hubs,Schneider+12,Peretto+14,Kumar+20,Kumar+22,Zhou+22_hubs}. Besides the morphological similarity with the astrophysical birthplaces of massive stars, these regions are indeed forming massive stars in the simulation. 

To further characterize them in relation to observed regions of high-mass star formation, we measure their mass-size relation within circular regions centered around the center of the maps, and containing 12.5\%, 25\% and 50\% of their total mass. The three circles are shown in all the maps of Figure~\ref{images_small}. The mass-size relation is plotted in Figure~\ref{mass_size} (squared symbols). These circular portions of the maps contain on average nearly 1,000\,$M_{\odot}$ at a scale of $\sim$0.5\,pc, or a mean column density of $\sim$0.1\,g/cm$^2$. The column density tends to decrease slightly with increasing size, following approximately the slope of the empirical mass-size relation limit for high-mass star formation of \citet{Kauffmann+2010}, as revised by \citet{Dunham+11}. The masses are on average a factor of three above the empirical limit, showing that the regions selected from the simulations have mean column densities consistent with those of observed regions forming high-mass stars. 

Figure~\ref{mass_size} also shows the mass-size relation of the brightest ATLASGAL regions (small, black circles) from \citet{Csengeri+2017}, that are supposedly some of the most extreme high-mass star-forming regions in the Galaxy, and that of the 15 regions chosen for the ALMA-IMF Large Program (blue circles) of \citet{Motte+22alma_imf}, which are a subsample of some of the brightest regions in \citet{Csengeri+2017}. The median value of the column density of the ATLASGAL clumps of \citet{Csengeri+2017} is a factor of 4.0 larger than that of our simulated regions, while the median column density of the even more extreme sample of the ALMA-IMF Large Program is 6.7 times larger. Nonetheless, there is partial overlap between these Galactic protocluster regions and our simulated ones, as the lowest mean column densities from both samples are well within the largest mean values from our maps (including the case of W43-MM3, shown by the lower of the two blue filled circles in Figure~\ref{mass_size}). The high mean {\em column} densities of the observed protocluster clumps may suggest very large mean {\em volume} densities, larger than in our simulation. However, as mentioned above and extensively discussed in \citet{Lu+2022_higal}, dust emission maps of regions in the Galactic plane and towards the Inner Galaxy may suffer important projection effects with contributions from dense structures at different distances. In general, the lower depth and column density of our simulation, compared with the most extreme regions in the Inner Galactic plane, should result in less projection artifacts and spatial confusion than in those regions. Thus, our modeling approach is conservative, in the sense that the issues raised in this work may be exacerbated in real observations of Galactic protoclusters.

\section{Synthetic Dust Continuum Maps} \label{synthetic_maps}

\subsection{Radiative Transfer} \label{radiative}

To generate the synthetic surface-brightness maps corresponding to the column-density maps of Figure~\ref{images_small}, we first carried out dust continuum radiative transfer modeling of the whole 250\,pc volume of each of the four snapshots with the continuum radiative transfer program SOC \citep{Juvela_2019}. Details of the radiative transfer method can be found in \citet{Lu+2022_higal}, and the main points are summarized below.

The radiative transfer calculations need the density field, the dust properties, and a description of the radiation field as inputs. The density field is taken from the datacubes of the AMR simulation, and SOC uses directly the same hierarchical spatial discretization as the AMR simulation. 
The dust properties correspond to the \citet{WeingartnerDraine2001} $R_{\rm V}=5.5$ dust model (Case B), where the size distribution of the carbonaceous grains extends up to 10\,$\mu$m. The high $R_{\rm V}$ is consistent with observations of very dense clouds, and the 1.3 mm dust opacity of the model, $\kappa_{1.3\,{\rm mm}}=0.0037\,{\rm cm^2\,g^{-1}}$, is more than twice the value of normal Milky Way dust. However, the opacity is still lower than for the \citet{Ossenkopf1994} dust model that is discussed for comparison in \S~\ref{temperature}, where $\kappa_{1.3\,{\rm mm}}=0.0083\,{\rm cm^2\,g^{-1}}$. The data for the \citet{WeingartnerDraine2001} dust models is obtained from the DustEm site\footnote{https://www.ias.u-psud.fr/DUSTEM/} \citep{Compiegne2011}.

The radiation field consists of an isotropic external field, the \citet{Mathis1983} model of the local interstellar radiation field, and of the radiation from all the stars with mass $>2M_{\odot}$ formed self-consistently in the simulation. The number of these point sources increases with time, as star formation progresses in the simulation, and is 775, 1929, 3093, and 3942 in the four snapshots of this work. The SOC code calculates the equilibrium dust temperature of each model cell and generates the corresponding 1.3\,mm surface-brightness maps for the three orthogonal view directions of each snapshot. These 250\,pc\,$\times$\,250\,pc surface-brightness maps have a pixel size equal to 0.0076\,pc, the smallest cell size of the simulation, hence they contain 32,768\,$\times$\,32,768 pixels. Of these maps, only the 2\,pc\,$\times$\,2\,pc regions corresponding to the selected column-density maps of Figure~\ref{images_small} are used for the analysis of this work.

\subsection{ALMA Simulations} \label{ALMA}

To generate the synthetic ALMA observations, the surface-brightness maps are processed with the {\sc CASA} program (v. 6.5.1), assuming a 2\,kpc source distance and using an ALMA 12-m antenna configuration ({\it alma.cycle8.1}) that gives a beam size of 1.66$\arcsec$ (geometric average for a source at zero declination). At equal linear resolution, this corresponds to a beam size of 0.60$\arcsec$ at the distance of 5.5\,kpc of W43, not far from the beam of 0.47$\arcsec$ in the study of W43 by \citet{Pouteau+22}. The {\sc CASA} runs include simulations with the {\em simobserve} routine and the cleaning of the dirty interferometer image with the {\em tclean} routine, using automatic masking (with the {\em auto-multithresh} option), Briggs weighting with the parameter {\em robust}=0, and multiscale deconvolution with the parameter {\em scales}=[0, 3, 9, 27]. The above steps are the same used in the study of W43 by \citet{Pouteau+22}, as described in \citet{Ginsburg+22}, except for the masking that was custom made. We do not need to distinguish between images generated from the full 1.3\,mm band and images that excludes the channels associated with lines \citep[referred to as {\em bsens} and {\em cleanest} images respectively in][]{Pouteau+22}, as our surface-brightness maps do not include any contamination from emission lines. The {\sc CASA} processing is carried out over an area of 3\,pc\,$\times$\,3\,pc containing the 2\,pc\,$\times$\,2\,pc region of interest, after doubling the number of pixels in the original surface-brightness maps, so the pixel size is approximately 0.4$\arcsec$, one fourth of the beam size.

   \begin{figure}
   \centering
   \includegraphics[width=\hsize]{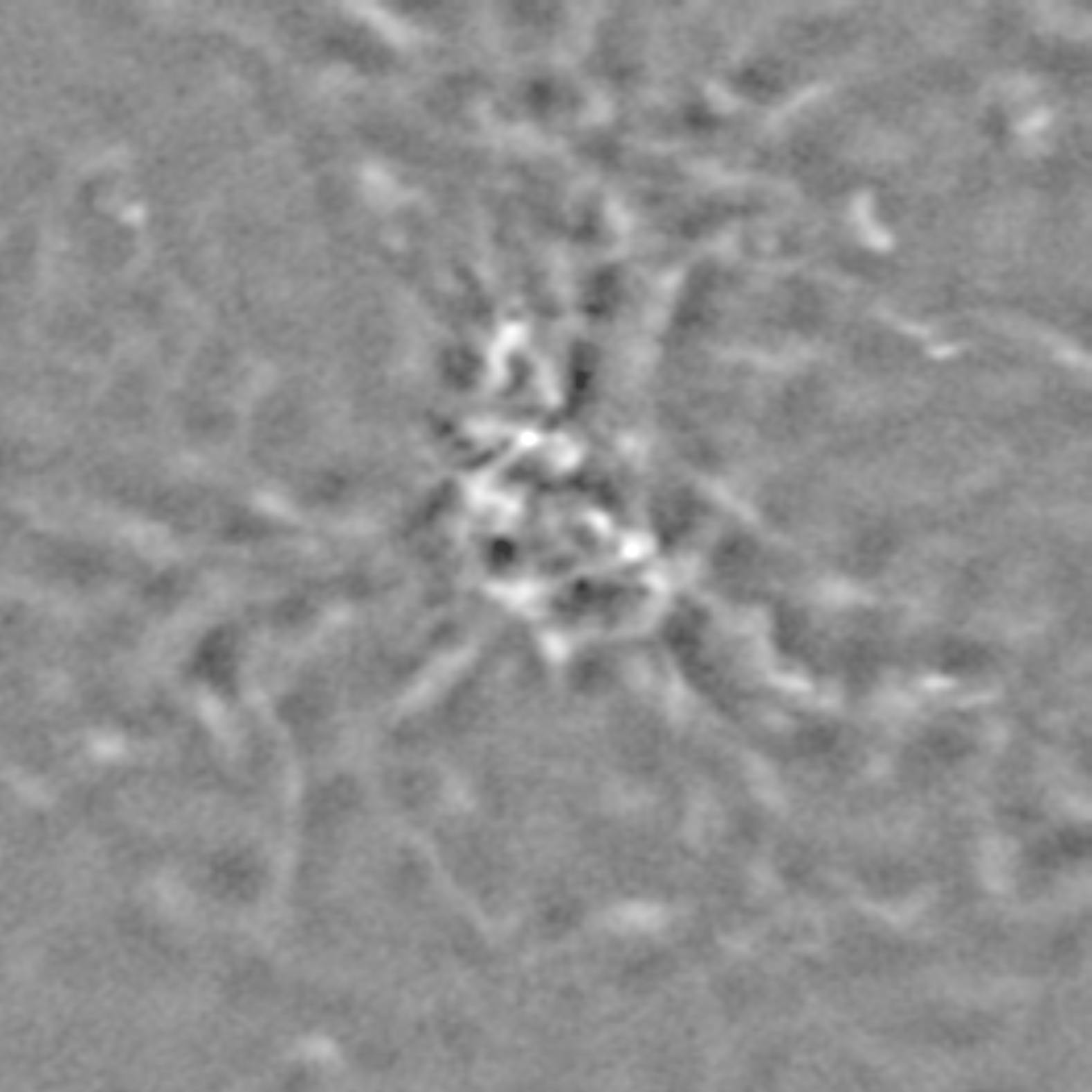}
   \includegraphics[width=\hsize]{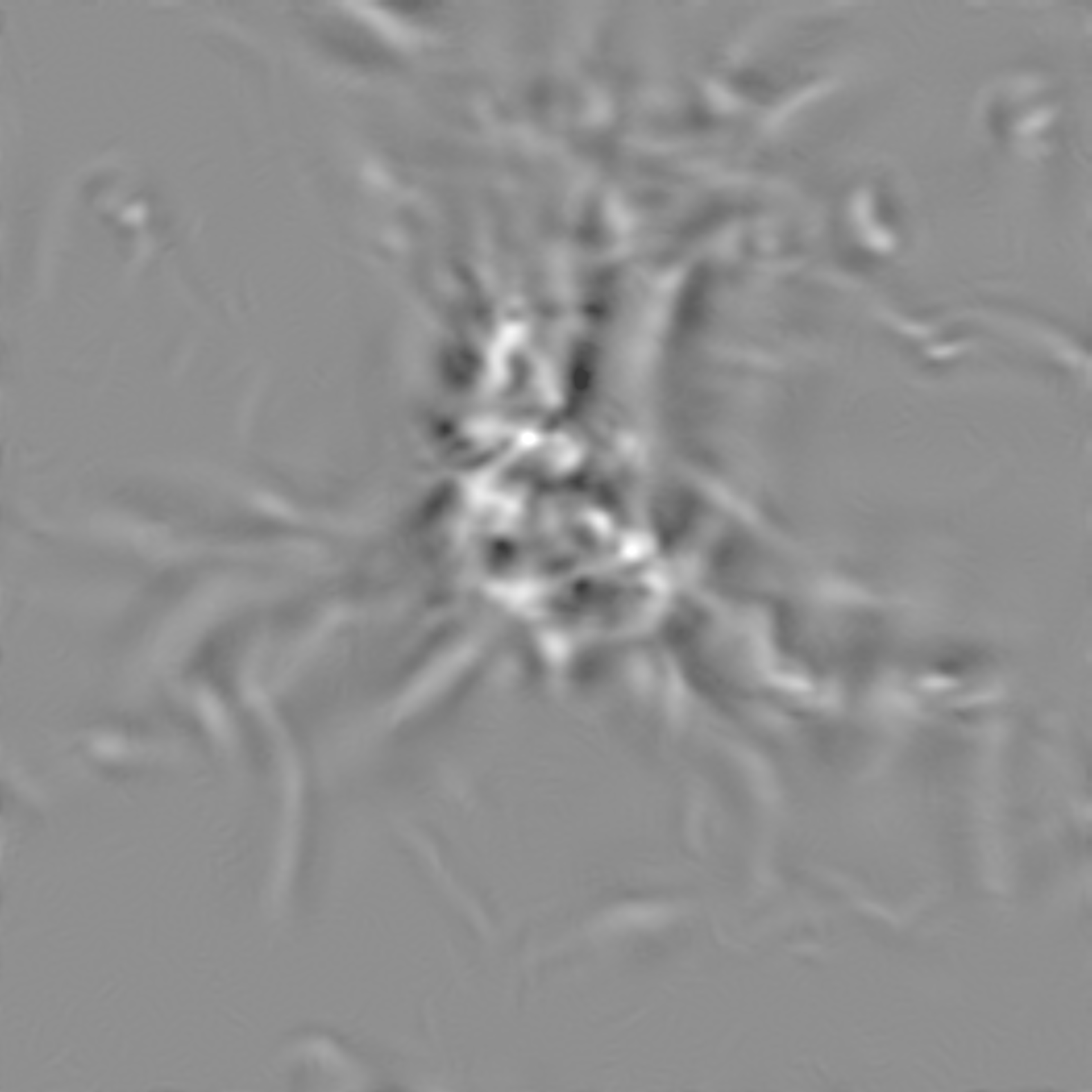}
      \caption{Example of the effect of noise reduction by the {\em MnGSeg} code on the map corresponding to the panel $3,z$ of Figure~\ref{images_small}. The upper panel is the simulated ALMA image before the noise reduction, the lower panel the image processed with {\em MnGSeg}. The grey scale is proportional to the surface brightness, with minimum and maximum values set to 5.0~MJy~sr$^{-1}$ (black colour) and 100.0~MJy~sr$^{-1}$ (white colour) respectively.
              }
         \label{denoising}
   \end{figure}  

We seek to obtain a similar sensitivity as in the ALMA-IMF Large Program that was designed to reach a point-source mass sensitivity of 0.15\,$M_{\odot}$ at 1.3\,mm (based on $T_{\rm dust}=20$\,K and $\kappa_{1.3\,{\rm mm}}=0.01\,{\rm cm^2\,g^{-1}}$). In the case of W43-MM2 and W43-MM3, at a distance of 5.5\,kpc, the noise level is approximately 0.06\,mJy\,beam$^{-1}$, based on Table~2 in \citet{Motte+22alma_imf}.\footnote{The estimated noise level is slightly larger in Table~1 of \citet{Pouteau+22}, probably due to the inclusion of regions of non-negligible dust emission.} With a 6\,hour integration time to obtain the whole 2\,pc\,$\times$\,2\,pc mosaic, the {\sc CASA} simulations with the parameter $PWV=0.5$ yield a noise level of $\sigma\approx 0.021$\,mJy\,beam$^{-1}$ at a distance of 5.5\,kpc, three times lower than in the maps of W43 of the ALMA-IMF Large Program. However, considering that our simulated regions have a mean column density that is on average 6.7 times smaller than that of the observed regions (see \S~\ref{protoclusters}), we further reduce the noise by averaging 8 different realizations of each map, where the {\sc CASA} simulations are exactly the same, except for the random thermal noise in the maps. After averaging the 8 maps, the typical noise level is $\sigma\approx 0.009$\,mJy\,beam$^{-1}$, estimated by measuring the rms value of the map in small regions with negligible column density. Thus, our point-source mass sensitivity is of order 0.3\,$M_{\odot}$, depending on the assumed dust temperature and opacity values.

\subsection{Noise Reduction} \label{noise}

Following the analysis in \citet{Pouteau+22}, we further reduce the noise in the maps with the multi-resolution segmentation code {\em MnGSeg} \citep{Robitaille+19}. {\em MnGSeg} separates the incoherent (Gaussian) cloud structures from the coherent component associated with the filaments and cores \citep[see][]{Robitaille+19}. We performed the separation using a value of 2.0 for the $q$ parameter and a skewness limit of 0.4. We then removed all the incoherent structures that were larger than the beam, as in \citet{Pouteau+22}, while those at scales equal to the beam or smaller were kept. This process reduces the noise and increases the number of detected sources by more than a factor of two. The effect of processing the images with {\em MnGSeg} is illustrated in Figure~\ref{denoising}, showing the synthetic ALMA images of the map $3,z$ of Figure~\ref{images_small} before (upper panel) and after (lower panel) the noise reduction.

\begin{figure}
   \centering
   \includegraphics[width=\hsize]{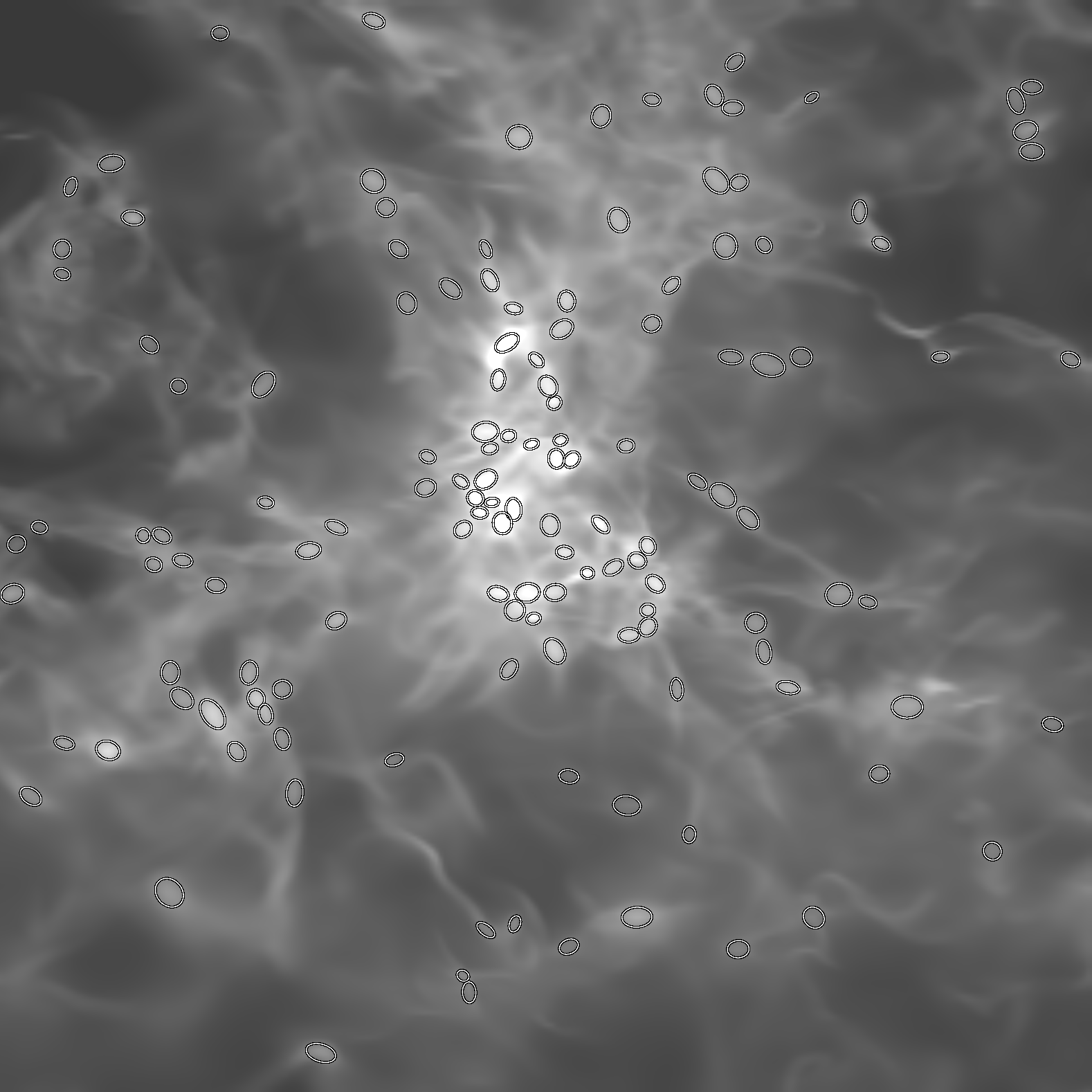}
   \includegraphics[width=\hsize]{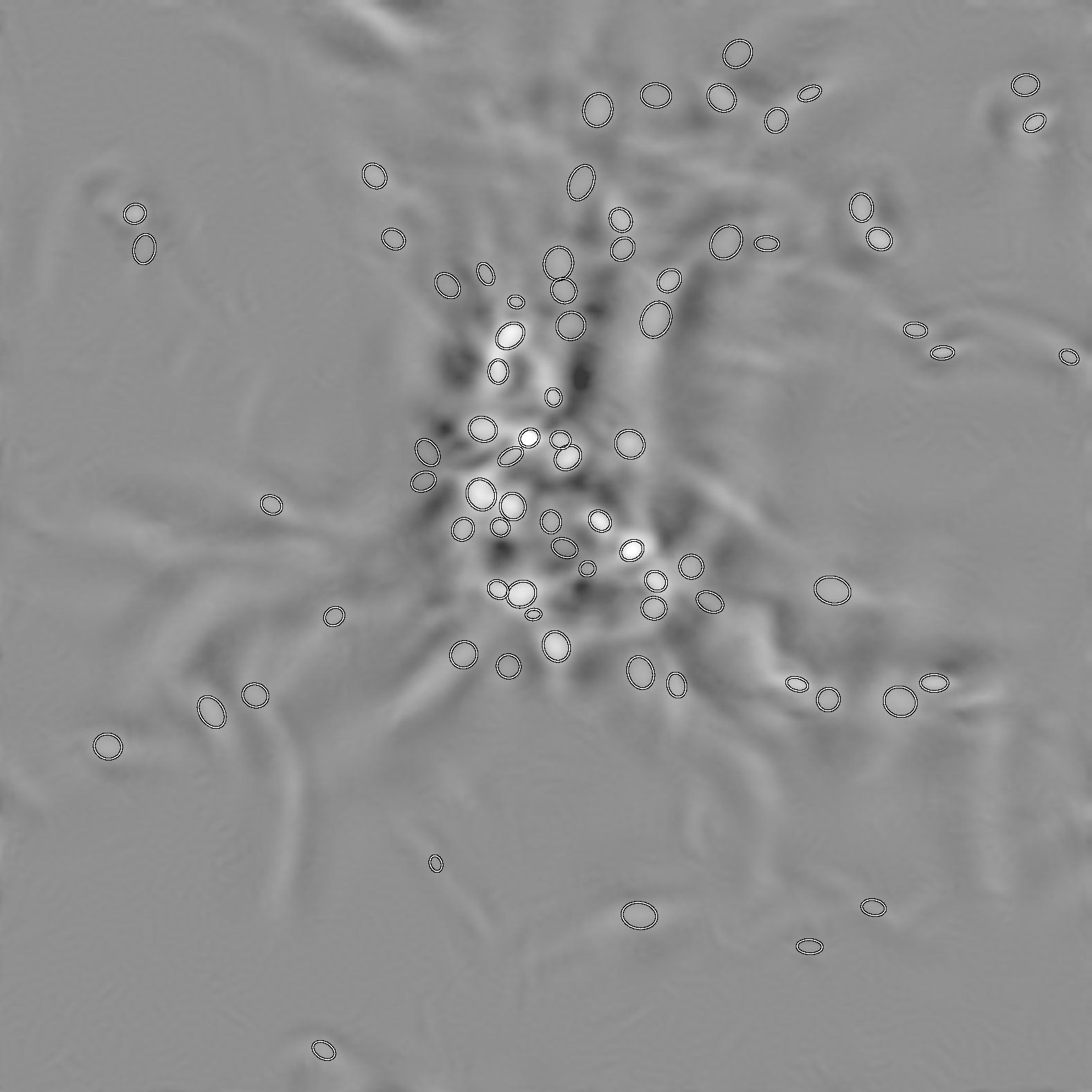}
      \caption{Examples of {\it getsf} core selection on a column density map (upper panel) and the corresponding synthetic ALMA map after noise reduction (lower panel), corresponding to the panel $3,z$ of Figure~\ref{images_small}. The grey scale is proportional to the square root of the column density (upper panel) or surface brightness (lower panel), with minimum and maximum values set to 0.05~g~cm$^{-2}$ (black colour) and 1.0~g~cm$^{-2}$ (white colour) in the upper panel and 5.0~MJy~sr$^{-1}$ (black colour) and 100.0~MJy~sr$^{-1}$ (white colour) in the lower panel. The ellipses correspond to the FWHM of the cores extracted by {\it getsf}.
              }
         \label{images_positive}
   \end{figure}


\section{2D and 3D Core Extraction} \label{getsf_clumpfind}

\subsection{2D {\em getsf} Extraction} \label{getsf}

The core extraction from both the column-density maps and the synthetic ALMA maps is carried out with the {\sl getsf} code \citep{Menshchikov21}, one of the two methods used in \citet{Pouteau+22}. Although several code parameters can be controlled by the user, all core (and filament) detection parameters have already been optimized and extensively tested \citep{Menshchikov21a,Menshchikov21}, and we adopt the standard extraction parameters as in \citet{Pouteau+22}, such as a maximum core ellipticity of 2 and a minimum signal to noise of 2. 

For each extracted core, {\sl getsf} determines a FWHM ellipse and computes the total mass, $M_{\rm c,2D}$, in the case of column-density maps, or the total flux, $S^{\rm int}_{\rm 1.3mm}$, from the surface-brightness maps. In the case of the synthetic ALMA maps, we derive the core masses from the {\sl getsf} integrated flux assuming the 1.3\,mm emission is optically thin: 
\begin{equation}
M_{\rm c,2D} = \frac{ S^{\rm int}_{\rm 1.3mm} \, d^2 } { \kappa_{\rm 1.3mm} \, B_{\rm 1.3mm}(T_{\rm dust}) }, 
\label{mass}
\end{equation}
where $d$ is the distance to the source, $\kappa_{\rm 1.3mm}$ is the dust opacity, $B_{\rm 1.3mm}$ is Planck's law, and $T_{\rm dust}$ is the dust temperature \citep{Hildebrand+77}. Although \citet{Pouteau+22} include a correction for optically thick emission \citep{Motte+18nature}, they find that the correction is significant for only two of their cores (with a change in the estimated mass of only 15\%), and we have verified that in our core samples that correction is always negligible. 
The dust opacity values are $\kappa_{\rm 1.3mm}=0.0037$\,cm$^2$\,g$^{-1}$ for the dust model of \citet{WeingartnerDraine2001} and $\kappa_{\rm 1.3mm}=0.0083$\,cm$^2$\,g$^{-1}$ for the dust model of \citet{Ossenkopf1994}. For the temperature, we use a single value for all cores, $T_{\rm dust}=13.0$\,K for the dust model of \citet{WeingartnerDraine2001} and $T_{\rm dust}=7.5$\,K for the dust model of \citet{Ossenkopf1994}, corresponding to the median dust temperature of the candidate 3D cores, defined in \S~\ref{getsf_cores} as the main overdensity in the line of sight of each {\sl getsf} core. All the analysis and figures in this work are based on the synthetic ALMA images from the dust model of \citet{WeingartnerDraine2001}. A comparison between the two dust models is shown in \S~\ref{temperature}.

We do not use the individual temperatures of the candidate 3D cores because they are not observable quantities. The color temperature can be derived from observations at different wavelengths, but we find that it does not accurately reflect the real dust temperature variations from core to core (see \S~\ref{temperature}), so there is no advantage in using it. Although in \cite{Pouteau+22} the core masses are based on estimates of individual core temperatures, we adopt a single value of $T_{\rm dust}$ also to estimate the core masses from the observations of the W43 region in \S~\ref{W43}. By using a single value of $T_{\rm dust}$, the derived masses are proportional to the integrated fluxes with the same constant of proportionality for all cores (see Equation~\ref{mass}). Thus, we effectively compare the integrated flux distributions rather than the CMFs, avoiding the extra uncertainty arising from the temperature determination.     

FWHM ellipses of the cores extracted by {\sl getsf} from one of our column-density maps and the corresponding synthetic ALMA map are shown in the two panels of Figure~\ref{images_positive}, which correspond to the panel $3,z$ of Figure~\ref{images_small}. The cores appear to follow the densest filaments and concentrate in the densest areas where multiple filaments intersect, as usually found in real Herschel and ALMA observations. Some of the same cores can be recognized in both images, but there are also several cases of sources extracted from the column-density map that are not seen in the synthetic ALMA map and vice-versa.

\subsection{3D {\em Dendro} Extraction} \label{dendro}

To extract the 3D cores from the 2\,pc\,$\times$\,2\,pc\,$\times$\,250\,pc columns of each map, we use the clumpfind algorithm introduced in \citet{Padoan+07imf}. To avoid confusion with other clumpfind algorithm in the literature, we here refer to this algorithm as {\sl Dendro} for the first time. While the algorithm was designed to select gravitationally unstable cores \citep[see][]{Pelkonen+21}, it can also be used as a straightforward dendrogram \citep{Rosolowsky+08}, without imposing a gravitational instability condition while building the tree. In this case, cores are simply defined as connected overdensities that cannot be split into two or more overdensities of amplitude $\delta n / n > f$. {\sl Dendro} scans the density field with discrete density levels, each of amplitude $f$ relative to the previous one. Only the connected regions above each density level are retained. The final (unsplit) core of each branch is retained. Each core is assigned only the mass within the density isosurface that defines the core (below that density level the core would be merged with its next neighbor). 

The {\sl Dendro} algorithm depends only on three parameters: (1) the spacing of the discrete density levels, $f$, (2) the minimum density above which cores are selected, $n_{\rm min}$, and (3) the minimum number of cells per core. In principle, there is no need to define a minimum density, but in practice it speeds up the algorithm. In this work we adopt $n_{\rm min}=10^4$~cm$^{-3}$, which is appropriate for prestellar cores. The parameter $f$ is set by looking for numerical convergence of the mass distribution with decreasing values of $f$. We find that the mass distribution above approximately 1\,$M_{\odot}$ is stable in the range of values between $f=32\%$ and $f=2\%$\footnote{In the range of masses between 1 and 10\,$M_{\odot}$, the slopes of the mass distributions computed with $f=2\%$ and $f=8\%$ differ by only 4\%.}, while at the lowest masses the number of cores slightly increases with decreasing $f$, but with a clear tendency to converge at around $f=2\%$. However, these variations in the core mass functions do not cause significant variations in our comparison between 2D and 3D cores, so the qualitative conclusions of this work do not depend on the precise value of $f$. The plots in this work correspond to the {\sl Dendro} runs with $f=8\%$. The minimum number of cells is set to 27, to ensure that all the detected cores are at least minimally resolved in the simulation and to limit the minimum mass and size of 3D cores to values comparable to those of the 2D cores extracted by {\sl getsf}.

Because {\sl getsf} distinguishes between sources and filaments, sources cannot have an aspect ratio larger than two. {\sl Dendro} does not impose a limit on the aspect ratio of the cores. However, we have verified that core elongation is not a problem in comparing with the {\sl getsf} 2D cores. For each 3D core, we have measured the half-mass distance of all its cells from its center and the equivalent radius. For elongated cores and filaments, the half-mass distance can be much larger than the equivalent radius. However, we find that an insignificant number of the 3D {\sl Dendro} cores have a ratio of half-mass distance over equivalent radius larger than two.

\section{Results} \label{results}

We present separately the results of the source extraction from the column-density maps and from the synthetic ALMA maps. The idealized case of the column-density maps provides a test of projection effects in isolation, without the observational complications, such as radiative transfer effects, interferometric artifacts and instrumental noise. These observational effects are then addressed in the second part of this section where the analysis of the synthetic ALMA maps is presented.

\subsection{Cores from Column Density Maps} \label{core_from_density}

\subsubsection{{\sl getsf} Runs and Candidate 3D Cores} \label{getsf_cores}

Using {\sl getsf} we extract 1,534 compact sources from the 12 column density maps, convolved with a beam size corresponding to 0.60$\arcsec$ at the distance of 5.5\,kpc of W43. We want to test if these cores correspond to real 3D cores, and, if so, how their masses compare with the corresponding 3D ones. One could directly search for overlap between 3D cores and 2D ones by simply projecting the positions of the 3D cores on the 2D maps. However, we find it more instructive to go through the extra step of first identifying a candidate 3D core position along the line of sight of each 2D core. This extra step provides a test of the quality of the background subtraction by {\sl getsf}, even when a counterpart real 3D core is not found. The {\sl getsf} background subtraction turns out to be rather impressive, in the sense that real 3D density structures are discovered by {\sl getsf} even if the background is often $\sim 90$\% of the mass in the line of sight.     

To identify a candidate 3D core position, we first search for the density maxima along the line of sight of each {\sl getsf} core, scanning its full 250\,pc length. The density is first averaged within the footprint of the core in the two directions on the plane of the sky, to obtain a 1D array of 32,768 elements, and a physical length of 250\,pc. Each element of this array corresponds to the mean density on a slice of thickness equal to 0.0076\,pc (the minimum cell size in the simulation) and area equal to the area of the core's footprint. The position of the highest density in this 1D array is chosen as the candidate 3D core position. We also define a candidate 3D core, as the spherical region centered around that position in the line of sight, and with radius equal to the footprint radius of the corresponding {\sl getsf} core. We refer to the mass of this 3D candidate core as $M_{\rm cand,3D}$. We repeat the process also for the second and third highest density maxima along the line of sight, so these other density peaks can also be considered when searching for an associated 3D core (see \S~\ref{dendro_cores}).

\begin{figure}
   \centering
   \includegraphics[width=\hsize]{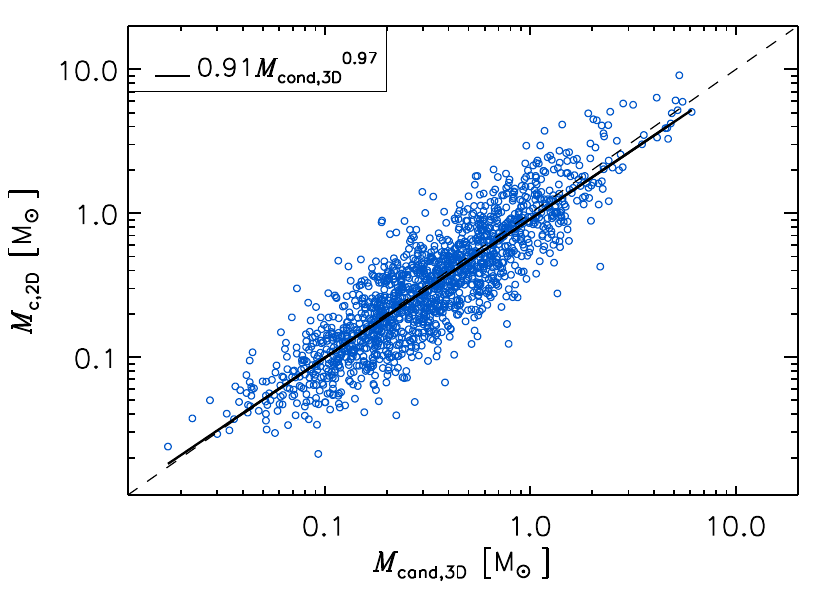}
   \includegraphics[width=\hsize]{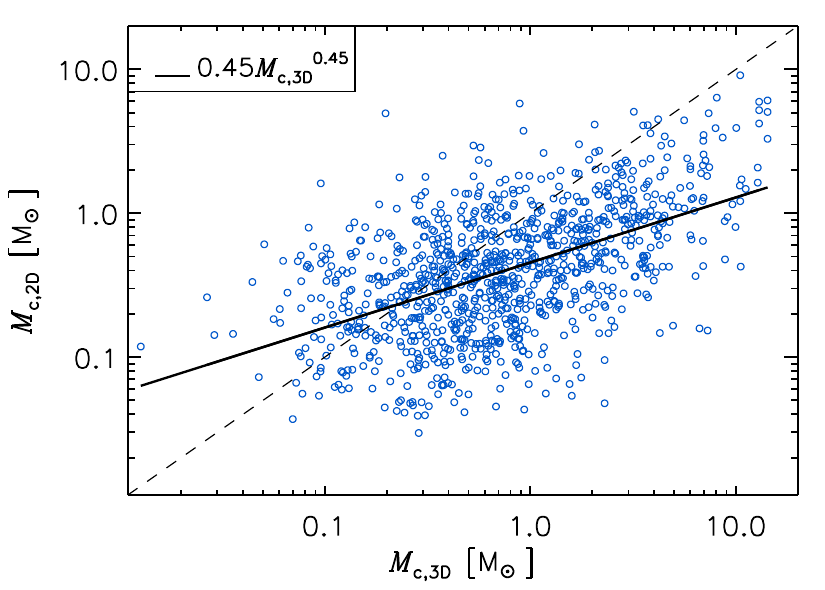}
      \caption{Comparison of the mass of cores selected by {\sl getsf} from the column-density maps, $M_{\rm c,2D}$, and the mass of the corresponding candidate 3D cores along the line of sight, $M_{\rm cand,3D}$ (upper panel), or the mass of the overlapping {\sl Dendro} cores extracted independently in the 3D data cube, $M_{\rm c,3D}$ (lower panel). The dashed lines correspond to equal masses, the thick solid lines to the least-squares fits. In the upper panel, all 1,534 {\sl getsf} cores are shown, while the lower panel is limited to the 1,023 overlapping {\sl getsf} and {\sl Dendro} cores.
              }
         \label{mass_2d_3d}
   \end{figure}

The comparison of the 2D {\sl getsf} masses with those of the corresponding candidate 3D cores is shown in the upper panel of Figure~\ref{mass_2d_3d}. The dashed line shows the one-to-one relation, while the thick solid line is a least-squares fit corresponding to the relation $M_{\rm c,2D}=0.91\,M_{\rm cand,3D}^{0.97}$ (where the mass unit is 1\,$M_{\odot}$). For any given value of $M_{\rm cand,3D}$, the error in the estimated mass, $M_{\rm 2D}$, is quite small. The standard deviation of $M_{\rm 2D}$ relative to the best fit in the logarithmic plot corresponds to a mass ratio of less than a factor of 1.3, the correlation is almost exactly linear, and the normalization at 1\,$M_{\odot}$ almost unity. Considering that the background subtraction performed by {\sl getsf} is on average quite large (approximately 90\% of the mass in the line of sight is interpreted as background in most cases), {\sl getsf} is very successful at extracting the mass of the dominant 3D dense structure along the line of sight. This also implies that the estimated volume density based on the {\sl getsf} mass and {\sl getsf} core size is also quite accurate, despite the complexity of the cloud morphology.


The success of {\sl getsf} to identify real high-density structures along the line of sight does not necessarily mean that most {\sl getsf} cores correspond to actual 3D cores of a similar mass, nor that most of the real 3D cores are captured in 2D by {\sl getsf}. There is no guarantee that the boundary of a candidate 3D core (a spherical volume of size equal to that of the {\sl getsf} core) corresponds to that of a core selected independently by a different 3D extraction algorithm. For example, a candidate 3D core may be a segment of a filament that happens to intersect another filament in projection (not in 3D), hence appearing as a bona fide 2D core. That filament segment with size equal to that of the {\sl getsf} core may indeed have a similar mass as that of the {\sl getsf} core, but it is not a real 3D core. Even when a candidate 3D core overlaps with a real 3D core, its mass (or that of the associated {\sl getsf} core) may not be a reliable estimate of that of the real 3D core, as the latter could have a significantly different size or center of mass. In addition, a majority of real 3D cores will generally be missed in 2D projections if they frequently overlap along the line of sight. Thus, to truly test the quality of the 2D source extraction relative to real 3D cores, we need to also independently extract 3D cores from the 3D datacubes that were used to generate the column-density maps. For that purpose, we use the {\sl Dendro} algorithm as described in \S~\ref{dendro}.

\subsubsection{{\sl Dendro} Runs and Real 3D Cores} \label{dendro_cores}

{\sl Dendro} returns a total of 7,298 3D cores (for $f=8\%$) from the 12 maps, a much larger number than found in 2D, as to be expected. The range of sizes and masses of the 3D cores are quite similar to those of the {\sl getsf} cores. The median mass is 0.25\,$M_{\odot}$, versus 0.27\,$M_{\odot}$ for the {\sl getsf} cores. We search for spatial overlap between the {\sl getsf} cores and the {\sl Dendro} cores by using the coordinates of the candidate 3D cores, as those cores are interpreted as the best 3D counterpart of the {\sl getsf} cores. If the center of a {\sl Dendro} core is found inside the volume of a candidate 3D core, then that {\sl Dendro} core is taken as the 3D counterpart to the corresponding {\sl getsf} core. If that condition is not satisfied, the test is repeated using the candidate cores associated to the second highest density peak in the line of sight, and failing also that, the third highest density peak is tested (see \S~\ref{getsf_cores}). We don't consider even lower density peaks below the third highest ones, because then the association with the {\sl Dendro} core is deemed to be insignificant (and indeed the correlation between the two 2D and 3D core masses would be negligible). Based on this criterion, we find a total of 1,023 overlapping cores, that is 67\% of the {\sl getsf} cores and 14\% of the {\sl Dendro} cores. 

The comparison of the masses of this subset of cores is shown in the lower panel of Figure~\ref{mass_2d_3d}. The least-squares fitting gives the relation $M_{\rm c,2D}=0.45\,M_{\rm c,3D}^{0.45}$ (with a mass unit of 1\,$M_{\odot}$). The vertical scatter in this plot is much larger than that in the upper panel, with a standard deviation relative to the best fit corresponding to a factor of four in mass. The fit is now significantly shallower than linear, and Pearson's correlation coefficient has decreased from 0.87 in the upper panel to 0.53 in the lower one. For each value of the 3D core mass, $M_{\rm c,3D}$, the corresponding mass of the overlapping {\sl getsf} core, $M_{\rm c,2D}$, spans a total range of nearly two orders of magnitude, not much smaller than the full range of {\sl getsf} core masses. Thus, the {\sl getsf} masses are rather unreliable estimates of the masses of the corresponding {\sl Dendro} cores, even if there is still a significant statistical correlation between the two masses.

\subsubsection{Core Mass Functions} \label{cmf}

\begin{figure}
   \centering
   \includegraphics[width=\hsize]{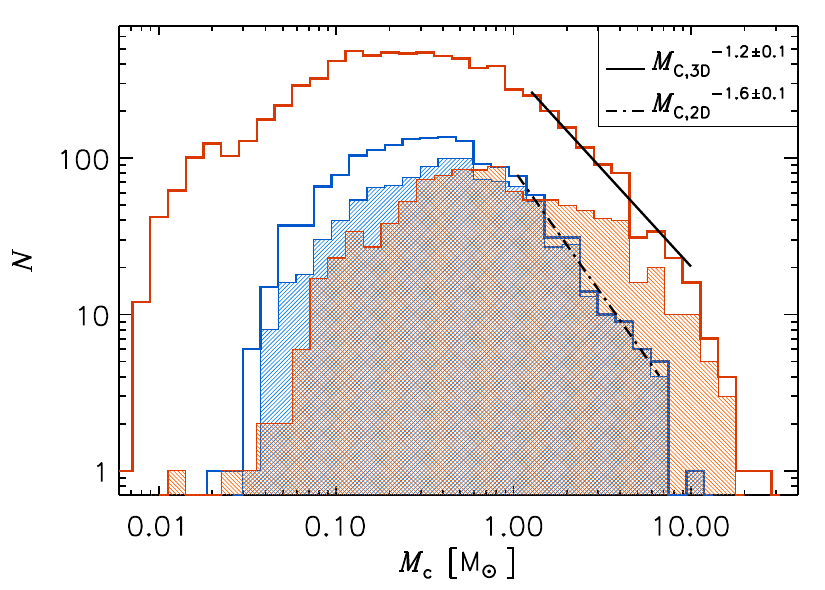}
      \caption{Mass distributions of the 3D {\sl Dendro} cores (red histograms), and of the 2D {\sl getsf} cores selected from the column-density maps (blue histograms). The unshaded histograms are for the full core samples, while the shaded ones are for the sub-samples of {\sl Dendro} cores (red histogram) and {\sl getsf} cores (blue histogram) that overlap with each other. Power-law fits in the approximate mass interval 1-10~$M_{\odot}$ are shown for the CMFs of the full samples.
              }
         \label{cmf_overlap}
   \end{figure}

Given that the number of {\sl Dendro} cores is so much larger than that of the {\sl getsf} cores, the great majority of the 3D cores (86\%) are not found in 2D, so the {\sl getsf} sample is highly incomplete. Figure~\ref{cmf_overlap} shows the CMFs of the 3D {\sl Dendro} cores (thick red line) and of the 2D {\sl getsf} cores (thick blue line). The number of 2D cores is always much smaller than the number of 3D cores, at any mass. The shape of the two CMFs are also somewhat different. In the range between approximately 1 and 10 $M_{\odot}$, the 2D CMF is significantly steeper, with a slope of $-1.6\pm 0.1$, than the 3D CMF that has a slope of $-1.2\pm 0.1$. The peak of the 3D CMF would certainly shift to lower masses if the numerical resolution of the simulation were increased \citep[e.g.][]{Pelkonen+21}, as more 3D cores of low mass would be found. However, assuming a fixed telescope beam size, the peak of the 2D CMF may not shift much towards lower masses even if the numerical resolution were increased, because the assumed telescope beam size in the {\it getsf} search is already twice larger than the current resolution of the simulation. 

\begin{figure}
   \centering
   \includegraphics[width=\hsize]{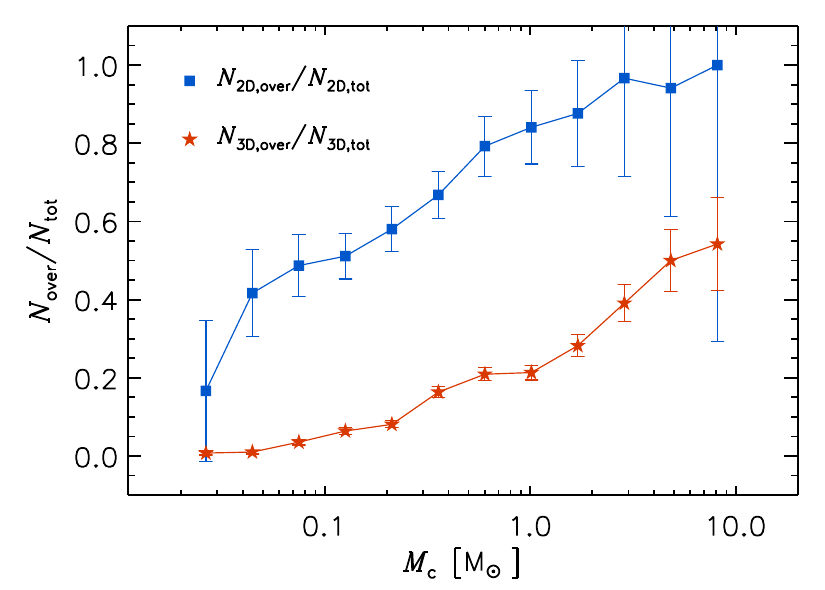}
      \caption{Fraction of cores with overlap between the {\sl Dendro} selection and the {\sl getsf} selection in the column-density maps, as a function of core mass. The blue squares are the fraction with respect to the total number of {\sl getsf} cores in each mass interval, $N_{\rm 2D,over}/N_{\rm 2D,tot}$, and the red stars the fraction with respect to the total number of {\sl Dendro} cores, $N_{\rm 3D,over}/N_{\rm 3D,tot}$. The error bars are the propagation of the Poisson errors of the individual quantities.
              }
         \label{overlap_fraction}
   \end{figure}

The CMFs of the 1,023 2D and 3D cores that are found to overlap with each other are shown by the shaded histograms in Figure~\ref{cmf_overlap}. The 3D CMF (red histogram) is clearly shifted to larger masses than the 2D CMF (blue histogram). The ratio of the 3D CMF of overlapping cores with the 3D CMF of all cores is shown by the red star symbols in Figure~\ref{overlap_fraction}. The plot shows that below approximately 0.5 $M_{\odot}$ less than 20\% of the 3D cores are found by {\sl getsf} in 2D, as clearly illustrated by the comparison of the shaded and unshaded red histograms in Figure~\ref{cmf_overlap}. The ratio increases monotonically with core mass, up to nearly 0.6 at the largest masses. 

Finally, the ratio of overlapping 2D cores and all 2D cores, shown by the blue squares in Figure~\ref{overlap_fraction}, gives the fraction of 2D cores that can be considered as real, rather than projection effects. This ratio is close to 1 above approximately 2 $M_{\odot}$, but decreases monotonically with decreasing core mass. At 0.1 $M_{\odot}$, approximately half of the cores are projection artifacts.

\subsection{Cores from Synthetic 1.3~mm ALMA Maps} \label{cores_from_ALMA}

\subsubsection{ALMA {\sl getsf} Runs and Candidate 3D Cores} \label{ALMA_getsf_cores}

We now consider the {\sl getsf} core extraction from the synthetic 1.3 mm ALMA maps corresponding to the column-density maps studied above. We refer to these 2D cores also as synthetic ALMA cores. The {\sl getsf} search yields a total of 791 synthetic ALMA cores, an average of 66 cores per map. The maps with the smallest and largest numbers of cores have 40 and 80 cores respectively. The reduced number of synthetic ALMA cores with respect to the number of cores from the column-density maps is to be expected, because the observational noise sets a finite mass sensitivity. 

Using the positions and sizes of the synthetic ALMA cores, we repeat the same study of density fluctuations in the line of sight of the {\sl getsf} cores as in the case of the column-density maps (see \S~\ref{getsf_cores}). The upper panel of Figure~\ref{mass_2d_3d_alma} compares the 2D {\sl getsf} masses with those of the corresponding candidate 3D cores, where the dashed line shows the one-to-one relation and the thick solid line is a least-squares fit. Although the vertical scatter is increased with respect to Figure~\ref{mass_2d_3d}, there is still a significant correlation between the two masses, with a Pearson's correlation coefficient of 0.64. However, the least-squares fit gives the relation $M_{\rm c,2D}=1.86\,M_{\rm cand,3D}^{0.66}$ (where the mass unit is 1\,$M_{\odot}$), much shallower than linear, and the {\sl getsf} masses are on average 2.3 times larger than those of the candidate 3D cores (based on the ratio of the median values), suggesting a smaller background subtraction than in the case of the column-density maps, possibly caused by interferometric artifacts, as discussed in \S~\ref{artifacts}. 

\subsubsection{ALMA {\sl Dendro} Runs and Real 3D Cores} \label{ALMA_dendro_cores}

\begin{figure}
   \centering
   \includegraphics[width=\hsize]{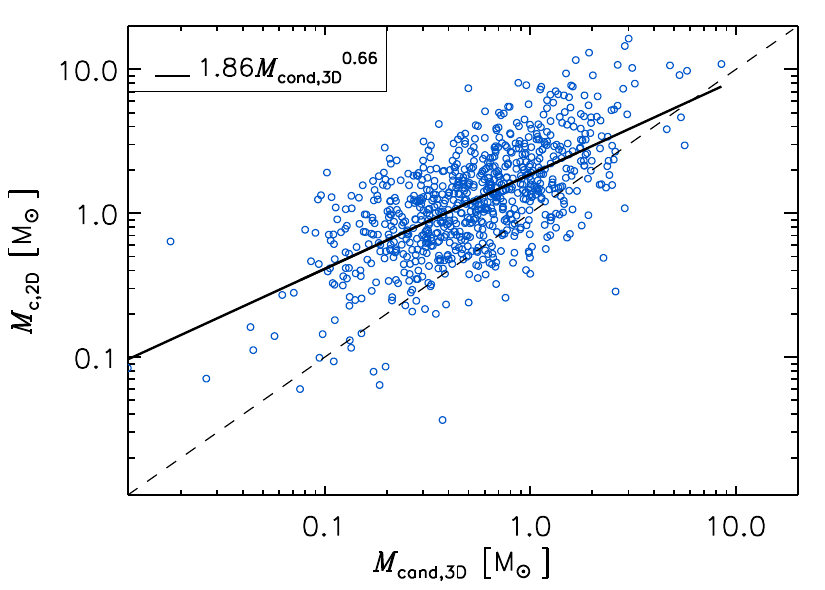}
   \includegraphics[width=\hsize]{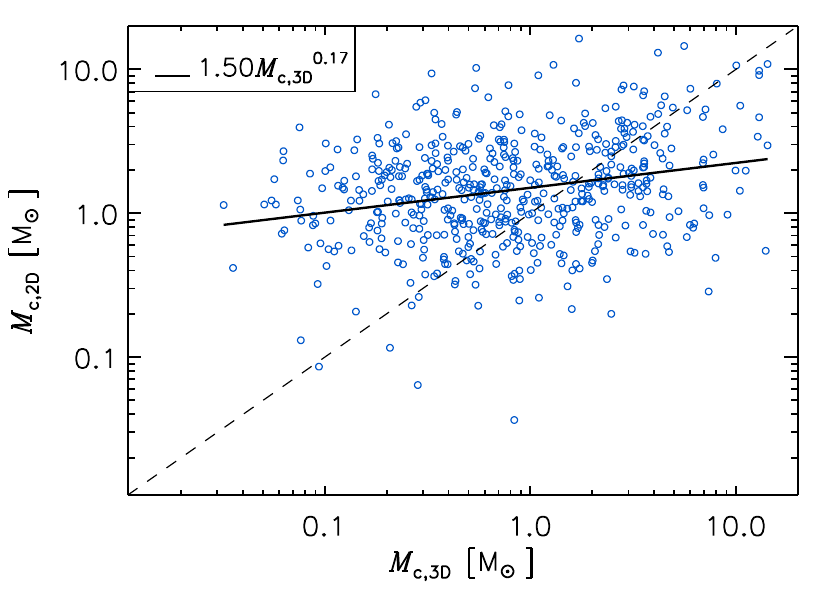}
      \caption{Comparison of the mass of cores selected by {\sl getsf} from the synthetic ALMA maps, $M_{\rm c,2D}$, and the mass of the corresponding main 3D cores along the line of sight, $M_{\rm main,3D}$ (upper panel), or the mass of the overlapping cores selected independently in the 3D data cube, $M_{\rm c,3D}$ (lower panel). The dashed lines correspond to equal masses, the thick solid lines to the least-squares fits. In the upper panel, all 791 {\sl getsf} cores are shown, while the lower panel is limited to the 557 {\sl getsf} cores that overlap with a corresponding 3D {\sl Dendro} core. 
              }
         \label{mass_2d_3d_alma}
   \end{figure}

Following the same procedure as in \S~\ref{dendro_cores} for the case of the column-density maps, we identify the 3D {\sl Dendro} cores that are associated with the synthetic ALMA cores extracted with {\sl getsf}. As in \S~\ref{dendro_cores}, the criterion is that the center of the {\sl Dendro} core must be inside the spherical volume of the candidate 3D core associated to the {\sl getsf} core. This search yields 557 overlapping cores, corresponding to 70\% of the synthetic ALMA cores, or 8\% of the 7,298 {\sl Dendro} cores. The lower panel of Figure~\ref{mass_2d_3d_alma} shows the comparison of the masses of the synthetic ALMA cores, $M_{\rm c,2D}$, with those of the corresponding 3D cores, $M_{\rm c,3D}$. The scatter in this plot is very large, and the two masses are barely correlated, with a Pearson's coefficient of 0.25, and an almost flat slope of the least-squares fit, $M_{\rm c,2D}=1.50\,M_{\rm c,3D}^{0.17}$ (where the mass unit is 1\,$M_{\odot}$). Despite the fact that 70\% of the synthetic ALMA cores have a real 3D counterpart (and approximately 90\% above 2\,$M_{\odot}$), their mass does not really reflect that of the corresponding 3D cores.  

We have verified that the correlation of the two masses remains rather low (Pearson's coefficient of 0.35) even if the {\sl getsf} masses were derived from the 1.3\,mm fluxes using the actual dust temperature of the candidate 3D cores (an ideal knowledge that could not be achieved from the observational data). Thus, the temperature uncertainty is only a partial contribution to the scatter, which is dominated by projection effects and by the differences between the 2D and 3D core definitions (as in the lower panel of Figure~\ref{mass_2d_3d}) and by interferometric artifacts (see \S~\ref{artifacts}). However, temperature variations along the line of sight certainly make projection effects and background subtraction more complex than in the case of the column density maps.

 \begin{figure}
   \centering
   \includegraphics[width=\hsize]{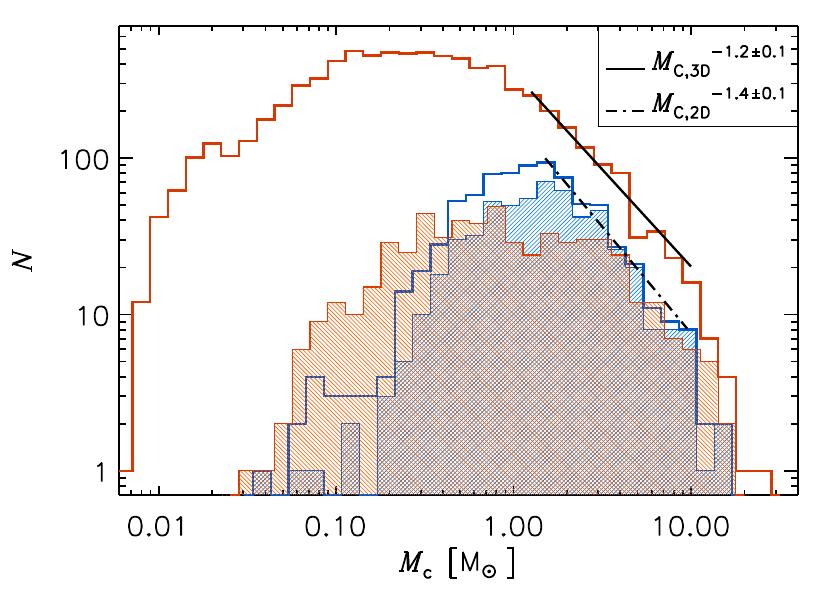}
      \caption{Mass distributions of the 3D-selected cores (red histogram), and of the {\sl getsf} cores selected from the synthetic ALMA maps (blue histogram). The unshaded histograms are for the full core samples, while the shaded ones are for the subsamples of 557   3D cores (red) and {\it getsf} ALMA cores (blue) that overlap with each other. Power-law fits in the approximate mass interval 1.5-10~$M_{\odot}$ are shown for the CMFs of the full samples.
              }
         \label{cmf_overlap_alma}
   \end{figure}

  \begin{figure}
   \centering
   \includegraphics[width=\hsize]{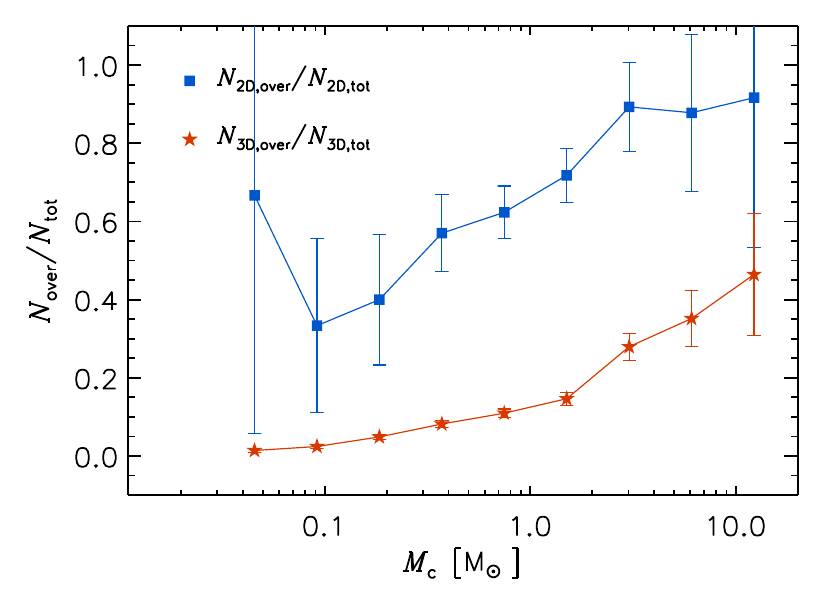}
      \caption{Fraction of cores with overlap between the 3D selection and the {\sl getsf} selection in the synthetic ALMA maps, as a function of core mass. The blue squares are the fraction with respect to the total number of {\sl getsf} cores in each mass interval, $N_{\rm 2D,over}/N_{\rm 2D,tot}$, and the red stars the fraction with respect to the total number of 3D cores, $N_{\rm 3D,over}/N_{\rm 3D,tot}$. The error bars are the propagation of the Poisson errors of the individual quantities.
              }
         \label{overlap_fraction_alma}
   \end{figure}

\subsubsection{ALMA Core Mass Functions} \label{ALMA_cmf}

The mass distribution of the synthetic ALMA cores is shown by the unshaded blue histogram in Figure~\ref{cmf_overlap_alma}. The CMFs of the 3D {\sl Dendro} cores is shown by the red unshaded histogram, while the shaded histograms show the CMFs of the subsamples of the 557 ALMA cores (blue) and 3D cores (red) that overlap with each other. In the range of masses where the CMF of the synthetic ALMA cores is close to a power law, approximately between 1.5 and 10~$M_{\odot}$, its slope is $-1.4\pm0.1$, significantly shallower than that of the {\sl getsf} cores from the column-density maps and only slightly steeper than that of the 3D {\sl Dendro} cores.  Above approximately 1~$M_{\odot}$, the number of synthetic ALMA cores is significantly larger than that of the 2D cores from the column-density maps (see Figure~\ref{cmf_overlap}), and only less than a factor of three lower than the number of 3D cores of the same mass. The reason for this excess of massive ALMA cores relative to those from the column-density maps may be related to non-trivial effects of interferometric artifacts, as discussed below in \S~\ref{artifacts}. 

The plot with blue-square symbols in Figure~\ref{overlap_fraction_alma} shows the ratio of the numbers of synthetic ALMA cores with a real 3D counterpart and the total number of synthetic ALMA cores, within each mass interval. Overall, 70\% of the synthetic ALMA cores have a 3D counterpart. Above 2~$M_{\odot}$, 88\% have a 3D counterpart or, conversely, only 12\% may be interpreted as projection artifacts. Towards lower masses, the fraction decreases monotonically (except for the smallest mass bin that is not statistically significant), down to a value of approximately 35\% at approximately 0.1\,$M_{\odot}$.    

The ratio between the 3D CMF of overlapping cores and the 3D CMF of all cores (shaded and unshaded red histograms in Figure~\ref{cmf_overlap_alma}) is shown by the red-star symbols in Figure~\ref{overlap_fraction_alma}. This ratio corresponds to the completeness fraction of the observational CMF. The fraction decreases monotonically with decreasing mass, as in Figure~\ref{overlap_fraction}, from 40\% at the largest masses, to 15\% at 1\,$M_{\odot}$, to 5\% or less below approximately 0.1\,$M_{\odot}$, near the empirical sensitivity limit of the synthetic ALMA maps. Evidently, the synthetic ALMA observations yield a highly incomplete sample of the real 3D cores. As to be intuitively expected, the completeness fraction decreases with decreasing mass at all masses. However, this does not result in a shallower 2D CMF relative to the 3D CMF at large masses, as trivially expected, because of the lack of correlation between 2D and 3D masses (Figure~\ref{mass_2d_3d_alma}). In other words, when comparing the blue and red unshaded histograms in Figure~\ref{cmf_overlap_alma},  the {\sl getsf} cores corresponding to the 3D {\sl Dendro} cores are not in the same mass bins in the two CMFs. This also means that the similar slopes of the two CMFs above approximately 1.5\,$M_{\odot}$ is essentially a coincidence: projection effects, 2D versus 3D core definitions, interferometric artifacts and radiative transfer effects (see \S~\ref{discussion}) transform the steep CMF of the 2D cores of the column density maps into a much shallower CMF of synthetic ALMA cores. Nevertheless, if the similarity in the slopes were confirmed as a general result, in practice one could use the observed slope as a rough estimate of the true slope of the CMF of 3D cores.

The continuous variation with mass of the completeness fraction of our synthetic core sample runs contrary to estimates for observational surveys of compact sources, where completeness fractions are usually of order unity above a minimum mass \citep[e.g.][]{Andre+10,Motte+18,Pouteau+22}, while it is consistent with what is found in the case of more extended cores \citep[e.g.][]{Cao+21_cygnus_CMF}. But we stress again that a simple correction of the CMF slope based on the mass-dependence of the completeness fraction is inadequate due to the lack of statistical correlation between 2D and 3D masses.

\section{Discussion} \label{discussion}

\subsection{Individual Effects} \label{individual}

We aim at evaluating the relative impact of different effects on the correlation between the {\sl getsf} ALMA core masses and the {\sl Dendro} 3D core masses shown in the bottom panel of Figure~\ref{mass_2d_3d_alma}. For that purpose, besides the column-density maps and the synthetic ALMA maps discussed in \S~\ref{core_from_density} and \S~\ref{cores_from_ALMA}, we also generate and analyze three more full sets of maps: single-dish, synthetic surface-brightness maps without thermal noise at 1.3\,mm with the same dust model \citep{WeingartnerDraine2001} and the same beam size ($0.60\,\arcsec$ at 5.5\,kpc) as the ALMA synthetic observations, synthetic ALMA images without thermal noise, and synthetic ALMA images with thermal noise, but without the noise reduction from the {\em MnGSeg} code. Synthetic cores are extracted with {\sl getsf} from each of these new sets of 12 maps, and the core masses are derived from Equation~\ref{mass} with the same $\kappa_{\rm 1.3\,mm}$ and $T_{\rm dust}$ values as for the synthetic ALMA cores (see \S~\ref{getsf}). 

\begin{figure}
   \centering
   \includegraphics[width=\hsize]{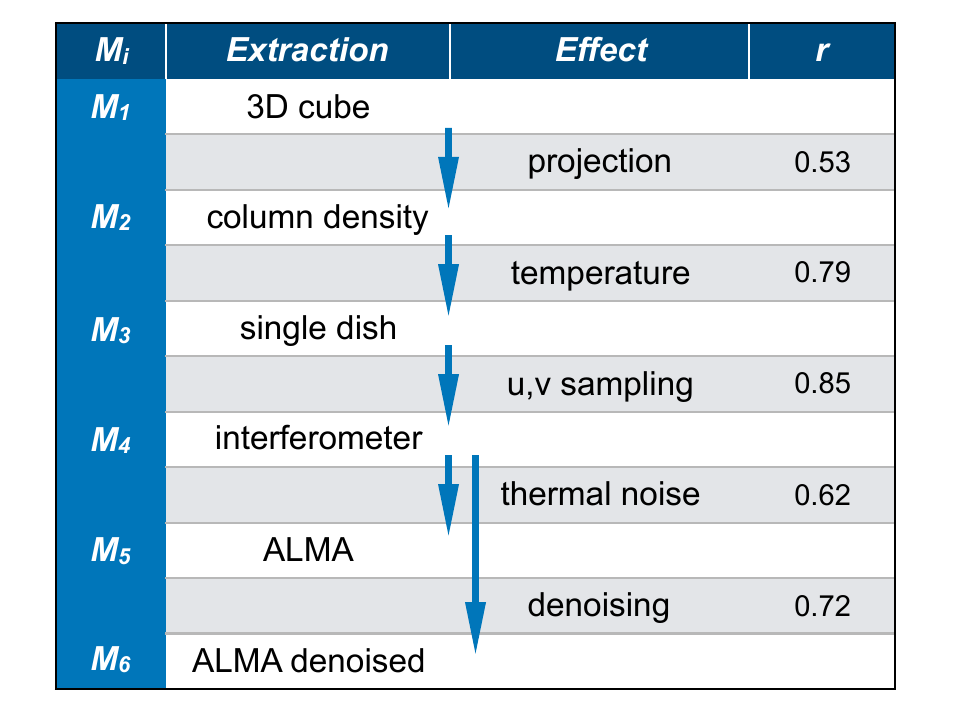}
      \caption{Schematic illustration of the different definitions of core masses, $M_{i}$, adopted in Figure~\ref{referee_multi}. The second column specifies the type of maps where the cores with masses $M_i$ are extracted from, the third column gives the main effect addressed by the comparison of core masses extracted in two consecutive ways, and the fourth column gives Pearson's correlation coefficient between the corresponding masses. `Single dish' stands for the beam-convolved surface brightness maps without noise, `interferometer' the simulated ALMA maps without noise, `ALMA' the simulated ALMA maps including thermal noise, and `ALMA denoised' the ALMA simulated maps with noise, where the noise has been reduced with the {\em MnGSeg} code.
               }
         \label{referee_table}
   \end{figure}

Including the {\sl Dendro} extractions from the 3D density cubes, we thus have six samples of cores, whose masses we refer to as $M_i$, with $i=1$ to 6, in the following. By comparing different core samples with each other, we can separate the contribution of different effects to the CMF. We consider these effects as consecutive steps in order of increasing complexity: projection effect (from the 3D cubes to the column density maps), temperature effect (from the column density to the single-dish surface brightness maps), (u,v) sampling effect (from the single-dish to the interferometric maps, without noise), noise effect (from the interferometric maps without thermal noise to the same maps with noise), and noise-reduction effect (from the interferometric maps without thermal noise to the same maps with noise first added, and then reduced with the {\em MnGSeg } code). We focus on the correlation of core masses between two consecutive steps in the order given above. These steps are summarized in Figure~\ref{referee_table}, where Pearson's correlation coefficient of each step is also given.

\begin{figure}
   \centering
   \includegraphics[width=\hsize]{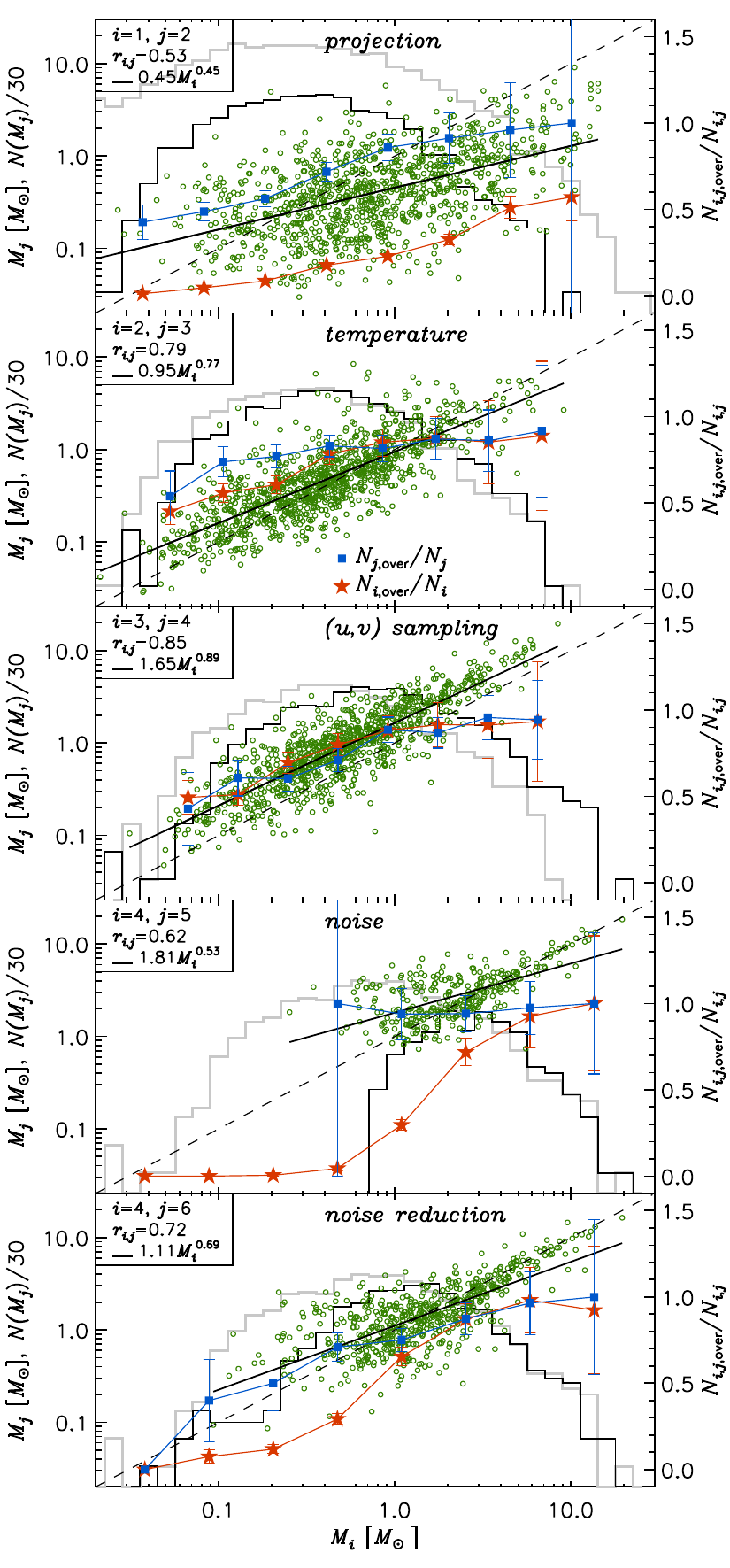}
      \caption{Scatter plots for masses of cores extracted in different ways, to highlight the different effects that contribute to the relation between the 3D {\sl Dendro} cores and the {\sl getsf} ALMA cores shown in Figure~\ref{mass_2d_3d_alma}. The definitions of the different masses, $M_{i}$, are given in Figure~\ref{referee_table} and further explained in the main text. The thick solid lines are the least-squares fits, whose results are reported in the legend, together with Pearson's correlation coefficients, $r_{i,j}$, for each pair of masses, $M_i$ and $M_j$. The light grey histograms show the mass distributions for the masses $M_i$, while the thick black histograms show those of the masses $M_j$. The blue squares are the fraction of $j$ cores that have an $i$ counterpart (so they are not artifacts with respect to the $i$ cores), and the red stars the fraction of $i$ cores with a $j$ counterpart (the completeness fraction), as in Figures~\ref{overlap_fraction} and \ref{overlap_fraction_alma}, with values given by the $y$ axis on the right hand side of the plots (notice that the $x$ axis is labeled as $M_i$, but in the case of the black histograms and the blue squares it should be $M_j$). 
              }
         \label{referee_multi}
   \end{figure}

Strictly speaking, the correlation between the initial masses $M_1$ (the 3D {\sl Dendro} cores) and the final masses $M_6$ (the {\sl getsf} ALMA cores) cannot be derived as the product of the individual steps. For example, the completeness fraction is not exactly the product of the individual completeness fractions, because cores that do not have a counterpart in the previous step may have a counterpart in the step before the previous one. However, we find that the product of the four correlation coefficients (excluding the case of noise without noise reduction that is shown here only as an alternative to the last step), is nearly identical to the correlation coefficient between $M_1$ and $M_6$: $r_{1,2}r_{2,3}r_{3,4}r_{4,6}=0.255$, while $r_{1,6}=0.251$, and the product of the individual completeness fractions is also a good approximation to the completeness fraction in Figure~\ref{overlap_fraction_alma}. Therefore, we are confident that this approach provides a good estimate of both the absolute and relative importance of each effect.  
The scatter plots between the pairs of masses $M_i$ and $M_j$ are shown in the five panels of Figure~\ref{referee_multi}, where the thick black lines are the least-squares fits with values shown in the legend. The green circles show the $M_i$ masses of cores with an $M_j$ counterpart (overlapping cores). The fraction of $i$ cores with a $j$ counterpart (completeness fraction), as a function of mass, is shown by the red stars, and the fraction of $j$ cores with an $i$ counterpart (interpreted as the fraction of real cores) is shown by the blue squares, as in Figures~\ref{overlap_fraction} and \ref{overlap_fraction_alma} (their values are given by the $y$ axis on the right hand side of the panels). Each panel also shows the CMFs for the $M_i$ masses (light grey histograms), and for the $M_j$ masses (thick black histograms)\footnote{For the grey histograms the $x$ axis label is correct, while the $y$ axis label should be $N(M_i)$, and for the black histograms (as well as the blue-squares plots) the $y$ axis labels are correct, while the $x$ axis label should be $M_j$.}.

The comparison of the five panels shows that the projection effect (top panel) has the strongest impact, meaning the lowest correlation coefficient as well as the lowest completeness fraction at large masses. It also generates a significant number of artifacts at low masses, as well as a steepening of the CMF, as already discussed in \S~\ref{dendro_cores}. Based on correlation coefficient, slope and completeness, the second strongest effect is noise. Comparing the two bottom panels, one can see that the noise reduction increases the correlation as well as the completeness at all masses, but at the price of a significant increase in the number of artifacts. While the case without noise reduction has essentially no artifacts, in the case with noise reduction the number of artifacts increases monotonically with decreasing mass, reaching approximately 30\% already at 1\,$M_{\odot}$. 

The effects of temperature and (u,v) sampling are comparable, both contributing to the CMF slope getting a bit shallower than that of the cores from the column-density maps. The (u,v) sampling effect is the main cause for the net shift of the CMF towards larger masses by approximately a factor of two, which is further discussed in \S~\ref{artifacts}. The contributions to the completeness and artifact fractions from the two effects are similar, with all four plots approximately monotonic across all mass values.


\begin{figure}
   \centering
   \includegraphics[width=\hsize]{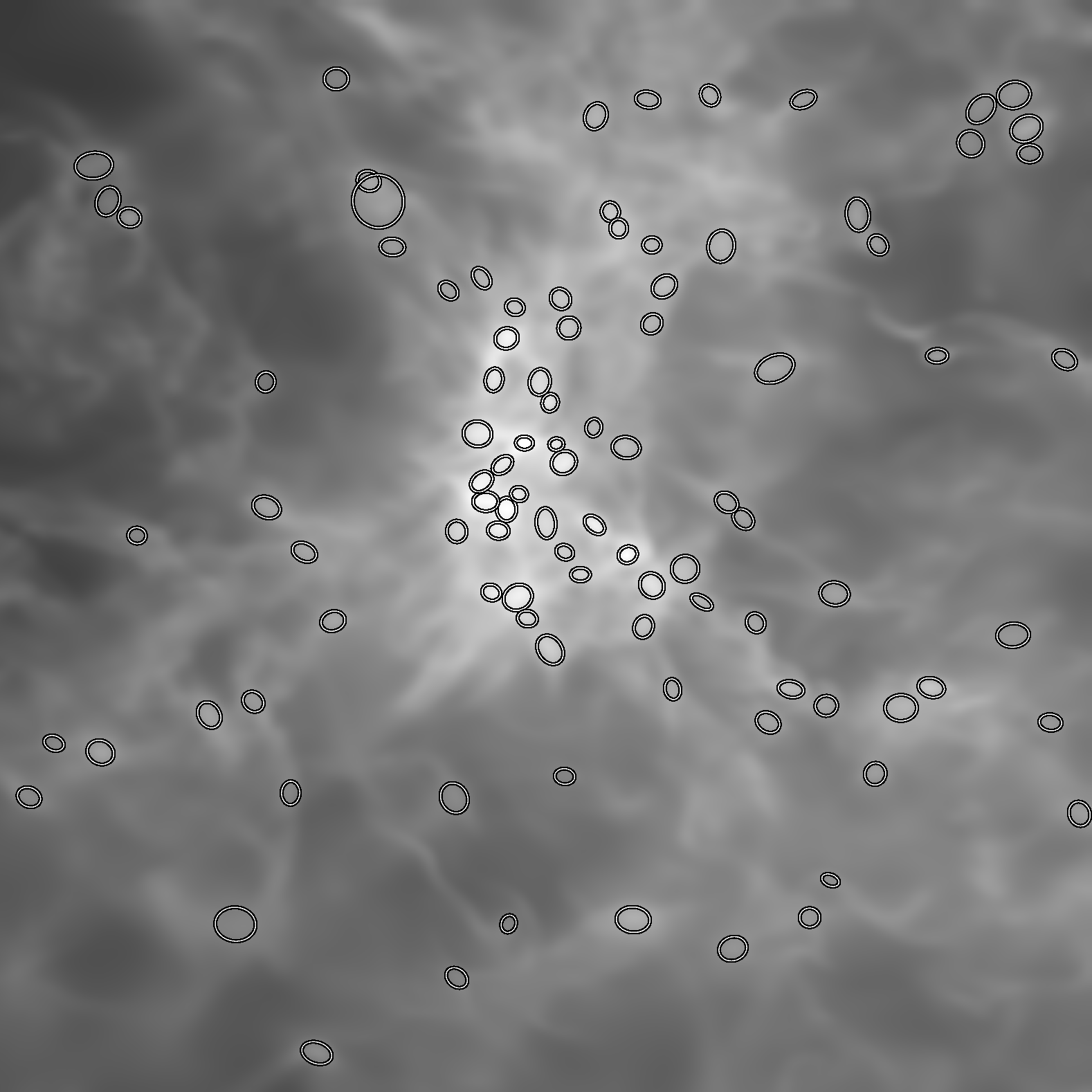}
   \includegraphics[width=\hsize]{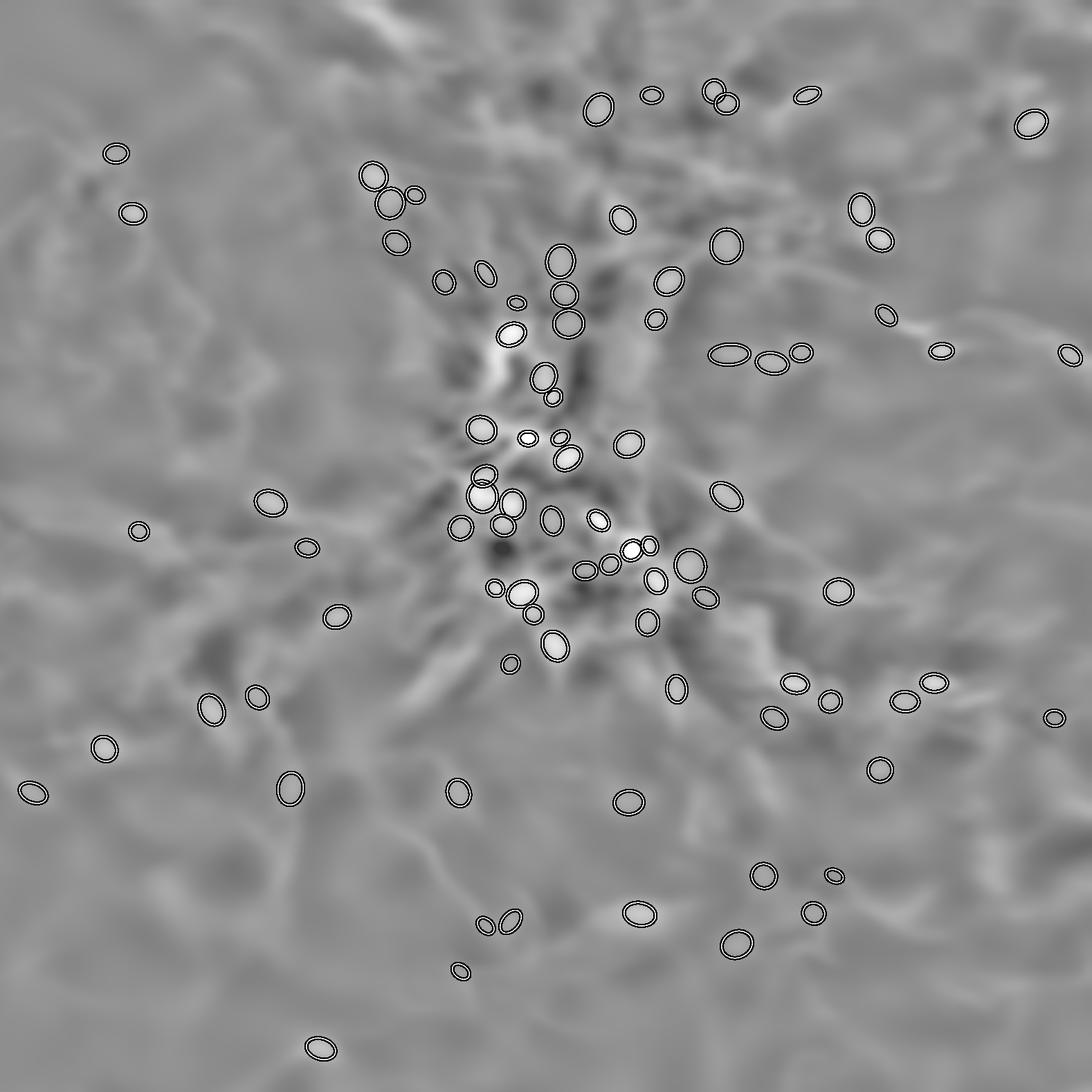}
      \caption{Examples of {\it getsf} core selection on a single-dish image without noise (upper panel) and the corresponding synthetic ALMA map with the same beam size and also without noise (lower panel), corresponding to the panel 3,z of Figure~\ref{images_small}. The grey scale is proportional to the square root of the surface brightness in both panels, with minimum and maximum values set to 5.0~MJy~sr$^{-1}$ (black colour) and 100.0~MJy~sr$^{-1}$ (white colour) in both panels. The ellipses correspond to the FWHM of the cores extracted by {\it getsf}.
              }
         \label{images_gauss_no_thermal}
   \end{figure}

  \begin{figure}
   \centering
   \includegraphics[width=\hsize]{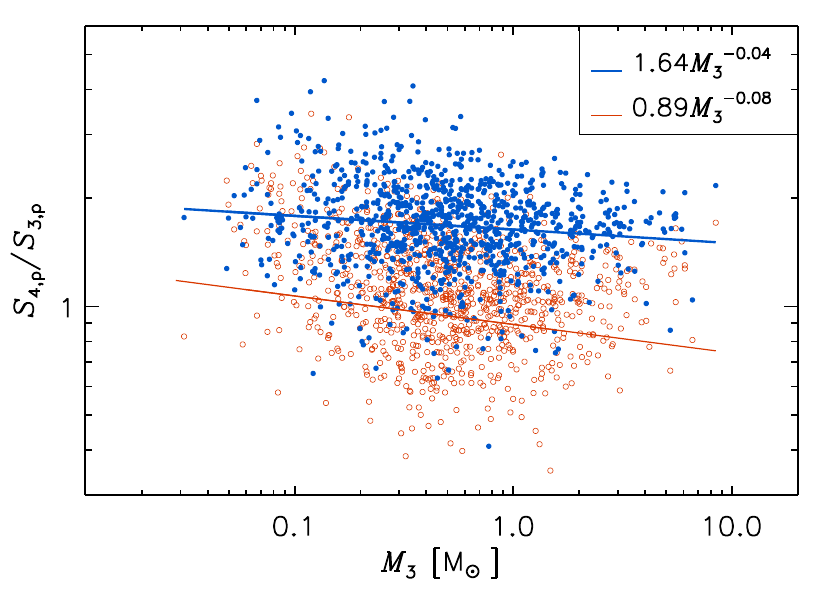}
      \caption{Ratio of peak surface brightness of {\sl getsf} cores selected from the simulated interferometric maps without noise, $S_{\rm 4,p}$, and from the simulated single-dish maps without noise, $S_{\rm 3,p}$. The blue, filled circles are for the extracted {\sl getsf} values, which include the background subtraction, while the empty red circles are for the values without background subtraction.  
               }
         \label{referee_background}
   \end{figure}

\subsection{Interferometric Artifacts} \label{artifacts}

We have shown in \S~\ref{ALMA_cmf} that the CMF of the synthetic ALMA cores has an excess of high-mass cores and a slightly shallower power-law tail relative to the CMF extracted from the column-density maps (blue unshaded histogram in Figure~\ref{cmf_overlap_alma} versus blue unshaded histogram in Figure~\ref{cmf_overlap}). The shift in the turnover mass depends on the noise level that sets a mass sensitivity: the higher the noise, the higher the turnover mass. But the high-mass tail of the CMF is also shifted to the right (larger masses). Two straightforward explanations for this excess of massive cores are possible: i) the observational noise causes an artificial merging of nearby cores extracted by {\sl getsf}; ii) the shift is due to an incorrect choice of the temperature (assumed to be too low). However, the comparison of the different CMFs in the panels of Figure~\ref{referee_multi} proves that neither of these simple explanations is correct.

The comparison of the two pairs of CMFs in the two bottom panels of Figure~\ref{referee_multi} shows that the effect of thermal noise is the expected one: the sensitivity of the images is reduced and so the CMF turnover shifts to larger masses, meaning that lower-mass cores are lost from the sample. But that loss does not cause an artificial merging into larger-mass cores, as the high-mass tails of the two CMFs from the maps with noise are indistinguishable from that of the CMF of the map without noise. Essentially the same high-mass cores are selected (or at least the same CMF of high-mass cores), so thermal noise is not the main cause of the shift to larger masses of the high-mass tail of the CMF of synthetic ALMA cores, relative to that of the cores extracted from the column-density maps. 

The second panel from the top of Figure~\ref{referee_multi} proves that radiative transfer effects, given the grain model and the corresponding core temperature adopted here, have a small effect on the CMF. So an incorrect choice of temperature is not the cause of the larger masses of the synthetic ALMA cores relative to those from the column density maps. However, if we had assumed a dust opacity inconsistent with that of the grain model used in the radiative transfer calculations, or an incorrect dust temperature of the cores, we would have obtained a significant shift in mass relative to the cores from the column-density maps. Thus, in the analysis of real observational data, the choice of dust opacity and core temperature will generally cause significant systematic mass uncertainties. In addition, if an individual temperature is estimated for each core, significant random uncertainties are also introduced. These considerations about dust opacity and dust temperature are further discussed in \S~\ref{W43} and \S~\ref{temperature}.    

The middle panel of Figure~\ref{referee_multi} shows that the main shift of the CMF towards larger masses occurs in the transition from the single-dish to the interferometric maps, where the same grain model and core temperature have been used. Thus, we can conclude that the high-mass tail of the synthetic ALMA CMF has an excess of massive cores relative to the CMF from the column-density maps, as well as a shallower slope, primarily due to interferometric artifacts (e.g. an increase of the luminosity contrast between the cores and their background). Because of the incomplete sampling of the (u,v) plane, interferometric images rely on interpolation whose weights are chosen based on assumptions about the source distribution. In the case of extended sources, one doesn't know a priori the real surface-brightness distribution, hence the incomplete (u,v) plane sampling and the interpolation process will always create artificial features in the final images, which are often referred to as interferometric artifacts (or interferometric noise).  

Figure~\ref{images_gauss_no_thermal} shows a comparison of {\it getsf} core selections on a single-dish image without noise (upper panel) and the corresponding synthetic ALMA map, also without thermal noise, with the same beam size (lower panel). Many cores in common between the two maps can be recognized, usually with comparable sizes. Yet we know that the cores in the lower panel are on average approximately a factor of two more massive than the cores in the upper panel. Although a detailed explanation requires a careful examination of the various extraction steps in {\sl getsf}, it is clear that the background subtraction is affected by interferometric artifacts creating some artificial intensity contrasts in the ALMA maps, resulting in a slightly less aggressive subtraction. Because the background subtraction is often of order 90\% of the column density, a small difference in background subtraction from 90\% to 80\%, for example, would cause an increase in the extracted core mass by a factor of two. This effect should be kept in mind, and possibly quantified, when assessing mass uncertainties of cores selected in interferometric surveys.

The scatter plots in Figure~\ref{referee_background} illustrate the effect of the background subtraction by comparing cores extracted from the single-dish maps with the corresponding cores extracted from the interferometric maps, both without noise. These are the same cores whose masses, $M_3$ and $M_4$ respectively, are plotted in the middle panel of Figure~\ref{referee_multi}. The blue filled circles in Figure~\ref{referee_background} give the ratio between the peak surface brightness of the cores from the two types of maps, $S_{\rm 4,p}/S_{\rm 3,p}$, as a function of the mass of the cores from the single-dish maps, $M_3$. The total scatter in the plot is approximately one order of magnitude, and the mean value of the peak-brightness ratio $\sim 2$, consistent with the scatter and mean values of the mass ratio, $M_4/M_3$, implied by the scatter plot in the middle panel of Figure~\ref{referee_multi}. However, the peak-brightness ratio without background subtraction is $\sim 1$, as shown by the empty red circles in Figure~\ref{referee_background}. This comparison of the two scatter plots, with and without background subtraction, confirms that the smaller background subtraction applied to the cores from the interferometric maps is the main reason for the shift of the CMF to larger masses, as discussed above.

\subsection{The CMF in the W43 Region} \label{W43}

  \begin{figure}
   \centering
   \includegraphics[width=\hsize]{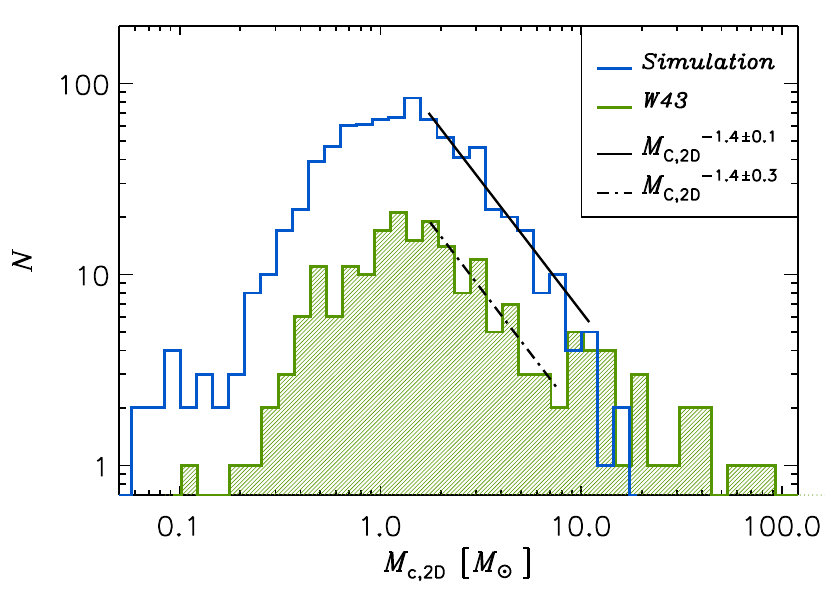}
      \caption{CMFs of {\sl getsf} cores selected from the synthetic ALMA maps (blue histogram) and of the W43-MM2 and MM3 cores from \citet{Pouteau+22} (green histogram) derived from the cores integrated fluxes and assuming a constant dust temperature (see text for details). Power-law fits in the approximate mass interval 2-10~$M_{\odot}$ are shown for both CMFs. 
              }
         \label{cmf_W43}
   \end{figure}


\citet{Pouteau+22} find that the high-mass tail of the CMF in the W43-MM2 and MM3 regions has a shallower slope than Salpeter's \citep[as in the case of W43-MM1, previously studied by][]{Motte+18}. This conclusion is based on plots of the cumulative CMF, which has the advantage of being independent of binning, but may also mask important features of the actual CMF. In the following comparison, we refer to the CMF, rather than its cumulative version. The comparison is independent of the uncertainty of the estimated core temperature, because all the core masses are derived from Equation~\ref{mass} using the same dust temperature value of $T_{\rm dust}=13.0$\,K. Thus, we effectively compare the distributions of integrated 1.3\,mm fluxes, rather than masses, even if the fluxes are expressed as masses. The mass conversion assumes $\kappa_{1.3\,{\rm mm}}=0.0037\,{\rm cm^2\,g^{-1}}$ for our synthetic cores (consistent with our dust model and previous plots), and $\kappa_{1.3\,{\rm mm}}=0.01\,{\rm cm^2\,g^{-1}}$ for the W43 cores, as assumed in \citet{Pouteau+22}. This choice of $T_{\rm dust}$ and $\kappa_{1.3\,{\rm mm}}$ results in a similar turnover mass for the CMF of W43 as for our synthetic cores. Using the median temperature estimated by \citet{Pouteau+22}, $T_{\rm dust}=23$\,K, the CMF of W43 would shift to lower masses by a factor of 0.6, but both mass scaling probably underestimate the true masses of the W43 cores by a factor of 2 to 4, depending on the dust models (see \S~\ref{temperature}). 

The CMF from W43 (green histogram) and of our synthetic ALMA cores (blue histogram), computed with a single temperature as mentioned above, are shown in Figure~\ref{cmf_W43}. In the limited range of masses between approximately 1.5 and 9\,$M_{\odot}$, where the CMF of W43 may be approximated by a single power law, its slope is $-1.4\pm 0.3$, consistent with that of our simulated regions. Both CMFs have a turnover at around 1.5\,$M_{\odot}$, and a similar drop towards lower masses, but the value of the turnover for the CMF of W43 is entirely dependent on the choice of $\kappa_{1.3\,{\rm mm}}$ and $T_{\rm dust}$, and would be shifted to larger masses if a combination of values consistent with either of our dust models were used (see \S~\ref{temperature}). The CMF of W43 shows a clear excess above approximately 9\,$M_{\odot}$, where 25 cores are found while only approximately 3 would be expected if the slope of -1.4 were to continue above 9\,$M_{\odot}$. 


  \begin{figure}
   \centering
   \includegraphics[width=\hsize]{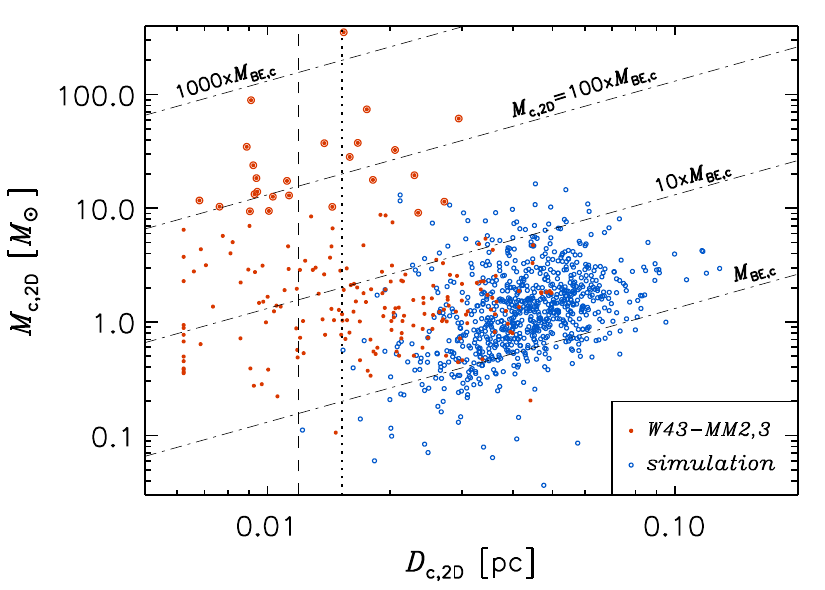}
      \caption{Mass-size relation for the synthetic ALMA cores (blue empty circles) from this work, and for the cores in the W43-MM2,MM3 region (red filled circles) from \citet{Pouteau+22}. The size, $D_{\rm c,2D}$, is the beam-deconvolved FWHM, and the mass, $M_{\rm c,2D}$, is derived from the 1.3\,mm flux assuming $T_{\rm dust}=13.0$\,K (see the main text for details). Larger red circles mark the W43 cores more massive than 9\,$M_{\odot}$. The diagonal dashed-dotted lines corresponds to constant values of $M_{\rm c,2D}/M_{\rm BE,c}$, where $M_{\rm BE,c}$ is the critical Bonnor-Ebert mass for a temperature of 13.0\,K. The vertical lines mark the beam sizes for the W43 observations (dashed line) and for the synthetic observations (dotted line). 
              }
         \label{mass_size_alma}
   \end{figure}

\subsection{Hypercritical Prestellar Cores, Massive Protostellar Cores, or Projection Effects?} \label{protostellar}

We can only speculate about the origin of the excess of cores above 9\,$M_{\odot}$ in the CMF of W43. To illustrate the special nature of these relatively massive cores, Figure~\ref{mass_size_alma} shows the mass-size relation of our synthetic ALMA cores with blue empty circles, and of the cores in W43 with red filled circles. The W43 cores above 9\,$M_{\odot}$ are marked by larger red circles. The size, $D_{\rm c,2D}$, is the beam-deconvolved FWHM, and the mass, $M_{\rm c,2D}$, is derived in the same way as for the CMFs of Figure~\ref{cmf_W43}. There is a significant overlap in the mass-size relation between our synthetic cores and those in W43, although the sizes of the synthetic cores are on average a factor of two larger. This is partly due to the fact that our beam size is slightly larger, $0.60\,\arcsec$ compared to $0.47\,\arcsec$ in W43 (see the dashed and dotted vertical lines in Figure~\ref{mass_size_alma}), partly a consequence of the limited resolution of the simulation (a bit too close to the beam size). In addition, a significant fraction of the W43 cores have FWHM size close to the beam size, so many of them are further reduced in size after beam deconvolution. As a result, 28\% of the W43 cores have deconvolved sizes smaller than the beam size, while only one of the synthetic cores has a deconvolved size smaller than the beam size.

\begin{figure}
   \centering
   \includegraphics[width=\hsize]{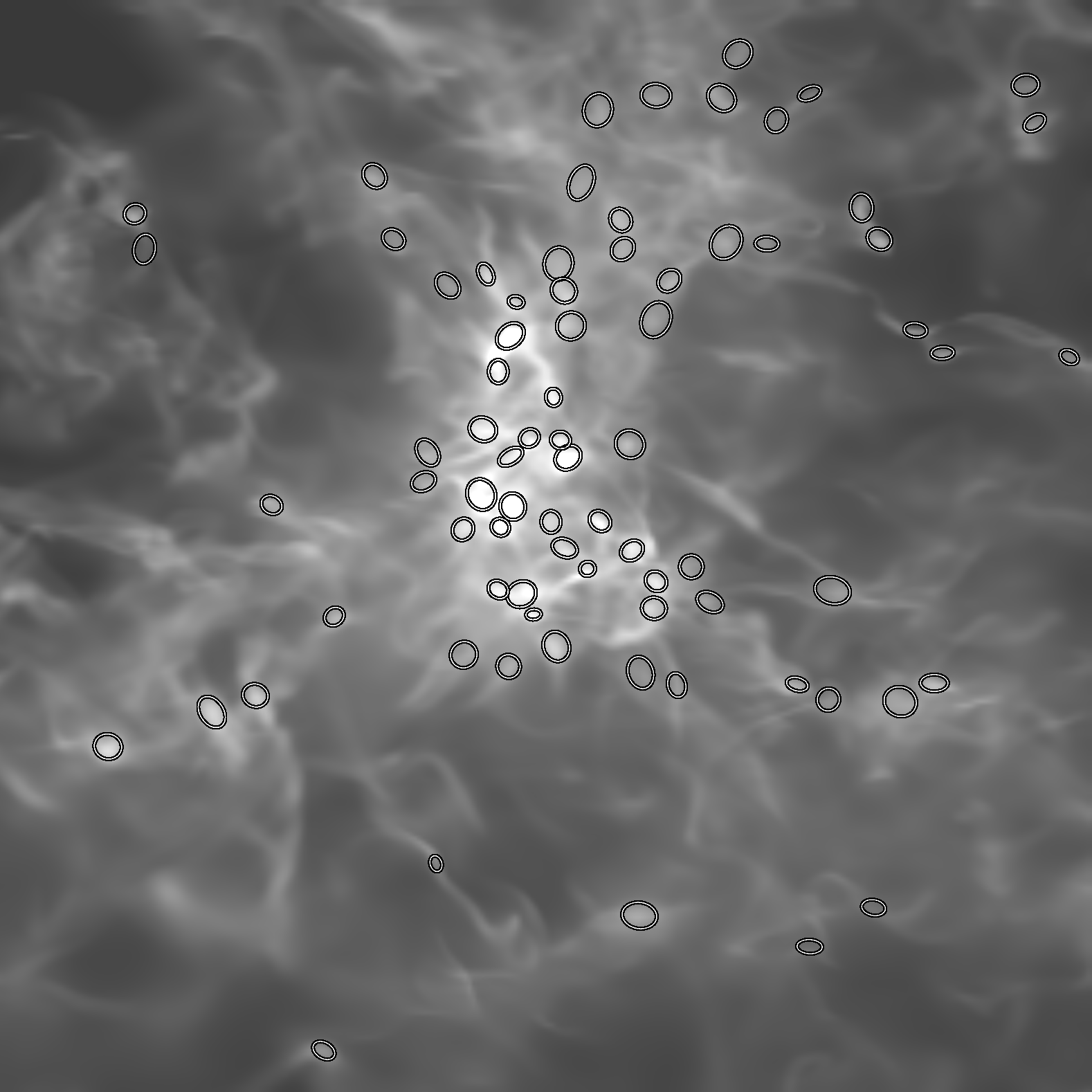}
   \includegraphics[width=\hsize]{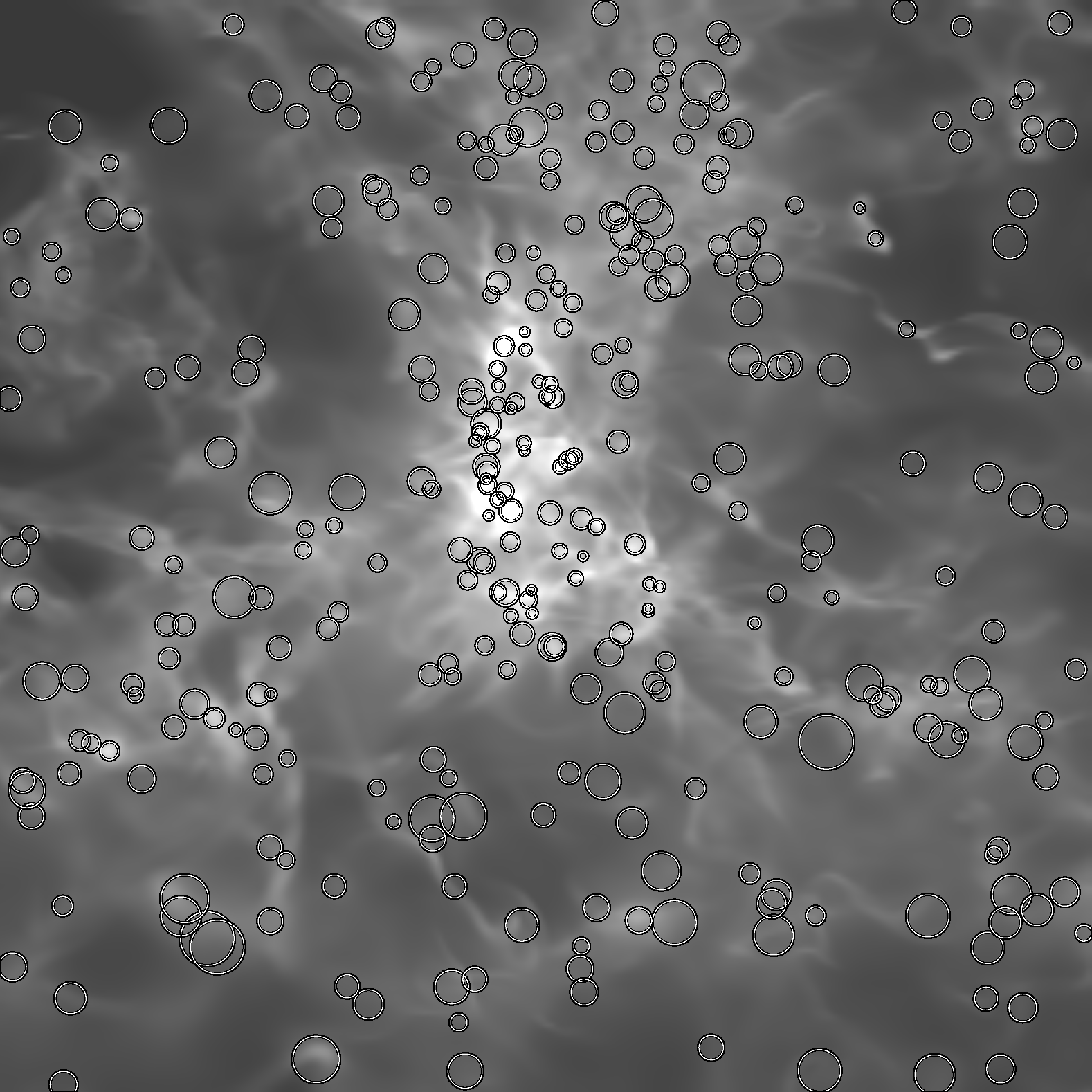}
      \caption{Comparison of 2D {\sl getsf} synthetic ALMA core positions (upper panel) with 3D {\sl Dendro} cores positions (lower panel) plotted over the same column-density image. The {\sl getsf} cores in the upper panel are the same as those in the lower panel of Figure~\ref{images_positive}. The lower panel shows the position and the size of the {\sl Dendro} cores with masses $> 0.5\,M_{\odot}$. Lower-mass {\sl Dendro} cores are omitted for clarity.
              }
         \label{images_2d_3d}
   \end{figure}

The diagonal dashed-dotted lines represent constant values of the ratio $M_{\rm c,2D}/M_{\rm BE,c}$, where $M_{\rm BE,c}$ is the critical Bonnor-Ebert mass \citep{bonnor56,Ebert57} for a temperature of 13.0\,K. Considering only the cores with masses below 9\,$M_{\odot}$, the median value of $M_{\rm c,2D}/M_{\rm BE,c}$ is 2.1 for the synthetic cores, and 5.1 for the W43 cores, due to the smaller sizes of the latter. The only four synthetic cores above 9\,$M_{\odot}$ have a median $M_{\rm c,2D}/M_{\rm BE,c}$ ratio of 20.5 (and a maximum value of 47.0), while the 25 cores in W43 have a median value of 118.6, with a maximum value of 1,747.\footnote{If pairs of values of $\kappa_{1.3\,{\rm mm}}$ and $T_{\rm dust}$ consistent with our dust models were used (see \S~\ref{temperature}), the $M_{\rm c,2D}/M_{\rm BE,c}$ ratios of the W43 cores would be even larger, by approximately a factor of three.} Thus, the W43 cores responsible for the CMF excess at high masses are all {\em hypercritical}, as their $M_{\rm c,2D}/M_{\rm BE,c}$ ratios are on average 50 times larger than those of the prestellar cores found in the simulation.     

It seems unlikely that such hypercritical cores are prestellar in nature, as in MHD simulations with realistic magnetic-field strength gravitational collapse initiates as soon as cores grow to be a few times their critical Bonnor-Ebert mass \citep[e.g. Figure~11 in][]{Padoan+20massive}. Their observed overabundance makes these cores even more unlikely to be prestellar, as the median density of the cores above 9\,$M_{\odot}$ in W43 is 43.0 times larger than that of the cores below 9\,$M_{\odot}$, so their free-fall time is 6.6 times shorter, making them very short lived.\footnote{The median value of $n_{\rm H_2}$ of the W43 cores above 9\,$M_{\odot}$ is approximately $8\times 10^7$\,cm$^{-3}$ (for the core mass contained within the FWHM size), which gives a free-fall time of only 3.4\,kyr.} Finally, if the core temperatures estimated in \citet{Pouteau+22} were correct for at least the most massive cores, they would imply a protostellar nature, as dust temperatures in excess of 15\,K are not expected at such high densities without internal heating sources with reasonable dust models (see \S~\ref{temperature} and references therein).

Thus, we speculate that, rather than being hypercritical prestellar cores, these objects are more likely to be protostellar cores. In that case, and given their small size, they may have already acquired a significant amount of rotation around the central star (depending on the magnetic-field strength and configuration), so they may be better described as pseudodisks. Even if this interpretation were correct, such compact protostellar cores could not be present in our simulation, because the accretion radius of the sink particles is four times the minimum cell size, or 0.03\,pc, so we cannot capture protostellar cores below a size of approximately 0.06\,pc (this also explains the low temperature of our cores, as they are essentially all prestellar). Future observations of emission lines at comparable angular resolution may confirm the protostellar nature of the massive cores in W43 through the presence of outflows, if not by direct evidence of infall or of supersonic rotation of the cores. Simulations at higher resolution will also be needed.\footnote{A new paper from the ALMA-IMF Large Program, \citet{Nony+23}, which appeared after the submission of this work, confirms our prediction that most of the massive cores in W43, responsible for the shallower CMF at large masses, are in fact protostellar in nature (up to 80\% for masses above 16\,$M_{\odot}$). Excluding the protostellar cores, \citet{Nony+23} find that the CMF of W43 has a slope of $-1.46{^{+0.12}_{-0.19}}$, consistent with our value of $-1.4\pm0.1$ for the CMFs of both W43 and our synthetic ALMA cores.} 

It is also possible that the mass estimate of a majority of the massive cores in W43 suffers from  projection effects and limited resolution and sensitivity. Projection effects must be even stronger in W43 than in our simulation, due to the higher maximum column density and the greater observational depth (several kpc) through the Galactic plane in the inner Galaxy. Figure~\ref{images_2d_3d} illustrates the difficulty of a 2D core extraction (upper panel) in a field with a large number of 3D cores (lower panel). The actual number of {\sl Dendro} cores is more than double of what is shown in the lower panel, as we only show the {\sl Dendro} cores with masses $> 0.5\,M_{\odot}$ for clarity. One can see that, even if our simulated region does not reach the column density where the most massive W43 cores are found, there is a significant chance of overlap of 3D cores in the densest regions, so a single {\sl getsf} core may have contributions from several {\sl Dendro} cores. Thus, it is possible that the mass of some of the most massive cores in W43 is overestimated due to the overlap of multiple 3D cores along the line of sight. These projection effects---overlap of real 3D cores along the line of sight---should not be confused with the artifacts mentioned in \S~\ref{results}, meaning 2D cores without a 3D counterpart. The case discussed here is quite the contrary: 2D cores with too many real 3D counterparts. However, if the most massive cores are protostellar, we cannot quantify the contribution of projection effects to their mass estimates, because protostellar cores are not captured by our simulation, as mentioned above.

Spatial resolution is also an important factor for compact-source extractions. When prestellar cores are selected as compact sources, they are not spatially resolved by definition. The largest ones, typically containing two to four beam sizes, have a measurable size, but their internal structure (e.g. the possibility of internal fragmentation) is unknown. Thus, cores extracted as compact sources should not be interpreted as individual objects in the absence of a strong physical justification for that assumption. The fact that cores extracted as compact sources are unresolved, in the sense that their sizes span a very limited range of values around the observational beam size, is well documented in the observational and numerical literature. Compact sources in the Hi-GAL Survey of the Galactic plane \citep{Elia+21}, extracted with the CuTEx code \citep{Molinari+11}, have sizes mostly below 2.5 times the (Herschel 250\,$\mu$m) beam size \citep[e.g.][]{Molinari+2016}. In the W43 regions MM2 and MM3, \citet{Pouteau+22} find only five {\sl getsf} sources with sizes larger than four times the beam size, which they exclude from their analysis. Almost 90\% of their sources have a size below 3.0 times the beam size, similarly to the Hi-GAL sources. Based on synthetic Herschel observations and the same data-analysis pipeline and source-extraction code as in the Hi-GAL survey, \citet{Lu+2022_higal} have shown that the Hi-GAL clumps are unresolved and their mass may not reflect at all that of the underlying cores (see their Section 7.3 and Figure~15). Using both Herschel and ALMA synthetic observations, \citet{Padoan+20massive} have shown that core masses may be highly overestimated, for example by a factor of 10 in the case of Herschel for regions at 1\,kpc distance (see their section 9.2 and Figures therein). They have also found that, even when the typical core masses are not overestimated on average, at the highest ALMA resolution, the estimated masses of individual cores show no correlation with the true core masses (see their Figure~27), in agreement with the results of this work \citep[considering that interferometric artifacts were not included in][]{Padoan+20massive}. A more straightforward study of {\sl getsf} core extractions in column-density maps from simulations (with no radiative transfer or interferometric effects, analogous to \S~\ref{getsf_cores} of this work) has demonstrated that the sizes of compact-sources depend on the beam size, hence the sources are unresolved, should not be interpreted as single objects, and beam deconvolution to extrapolate their size is not justified \citep{Louvet+21}.     

\begin{figure}
   \centering
   \includegraphics[width=\hsize]{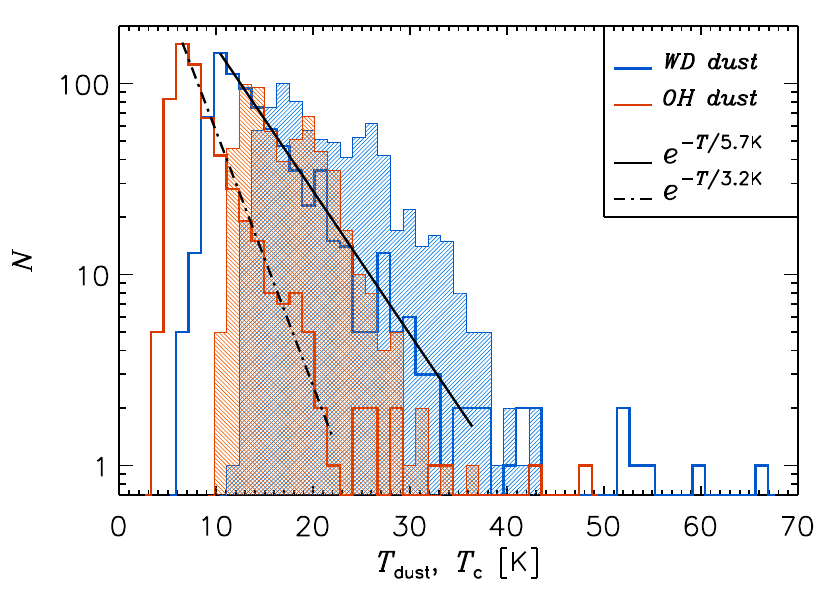}
      \caption{Distributions of the mass-weighted mean dust temperature, $T_{\rm dust}$, of the candidate 3D cores (defined in \S~\ref{getsf_cores}) for the WD dust model (blue unshaded histogram) and the OH dust model (red unshaded histogram). The black solid and dashed-dotted lines are the least-square fits reported in the legend. The shaded histograms are the distributions of the color temperature, $T_{\rm c}$, of the corresponding {\sl getsf} cores (see details in the main text) for the WD model (blue histogram) and the OH model (red histogram).    
              }
         \label{dust_models_histogram}
   \end{figure}

\subsection{Dust Models and Core Temperature} \label{temperature}

All previous plots in this work are based on the \citet{WeingartnerDraine2001} $R_{\rm V}=5.5$ dust model (Case B), with $\kappa_{1.3\,{\rm mm}}=0.0037\,{\rm cm^2\,g^{-1}}$. However, we have carried out the full analysis also with the dust model of \citet{Ossenkopf1994}, corresponding to dust that has evolved for $10^5$ years at a density of $10^6$\,cm$^{-3}$ and has acquired thin ice mantles. This model has a significantly larger dust opacity, $\kappa_{1.3\,{\rm mm}}=0.0083\,{\rm cm^2\,g^{-1}}$. Although the conclusions of this work are not affected by the choice of the dust model, here we provide a brief comparison between the two cases, as well as a discussion of core temperatures. In the following, we refer to the two dust models as the WD model and the OH model respectively. 

Figure~\ref{dust_models_histogram} shows the distribution of the dust temperature of the candidate 3D cores (see \S~\ref{getsf_cores}), computed as the mass-weighted mean temperature of the dust inside the spherical volume of each core. The unshaded blue histogram is for cores extracted from the synthetic ALMA maps based on the WD model, while the unshaded red histogram from the maps computed with the OH model. As expected from the larger dust opacity, the OH model results in lower dust temperature at the high densities found in the cores. The median dust-temperature values, used to derive the core masses (see \S~\ref{getsf}), are 13.0\,K for the WD model and 7.5\,K for the OH model. Both temperature distributions are very well approximated by exponential functions for dust temperatures above the peak (corresponding roughly to the median value), except for a few high-temperature outliers.

\begin{figure}
   \centering
   \includegraphics[width=\hsize]{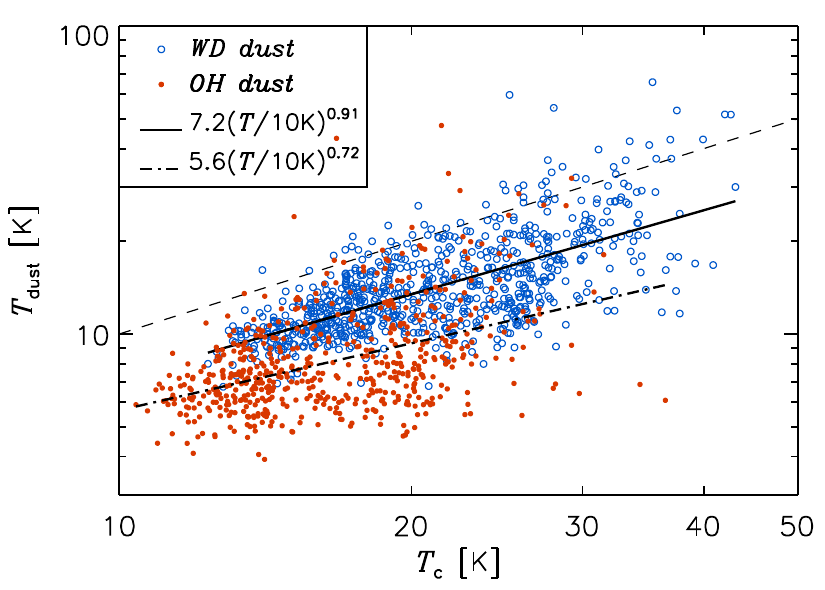}
      \caption{Dust temperature (of the candidate 3D cores) versus color temperature (of the corresponding {\sl getsf} cores) from the WD dust model (blue empty circles) and the OH dust model (filled red circles). The black solid and dashed-dotted lines are the linear fits reported in the legend, and the thin dashed line is the one-to-one relation.      
              }
    \label{dust_models_scatter}
   \end{figure}

\begin{figure}
   \centering
   \includegraphics[width=\hsize]{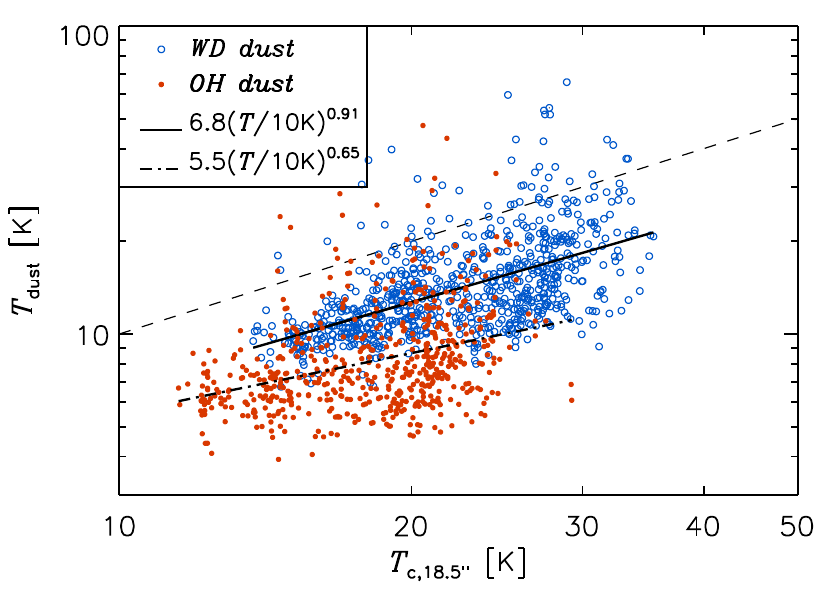}
      \caption{The same as Figure~\ref{dust_models_scatter}, but with $T_{\rm c}$ computed at an angular resolution of 18$\arcsec$.       
              }
    \label{dust_models_scatter_Herschel}
   \end{figure}

We also compute the color temperature, $T_{\rm c}$, of all the synthetic ALMA cores, using surface brightness maps at 100, 160, 250, 350, 500, 850, and 1300\,$\mu$m. The wavelengths are similar to those used in \citet{Pouteau+22} (apart from the exclusion of the 70\,$\mu$m band and the 870\,$\mu$m wavelength being replaced with 850\,$\mu$m). In the modified blackbody fits, we assume a constant dust opacity spectral index of $\beta$=1.8 and the same relative uncertainty at all frequencies. The final color temperature estimates are averages over the 2D footprint of each core. This method is less sophisticated than that used in \citet{Pouteau+22}, which is based on the PPMAP method of \citet{Marsh+15}, with some other corrections for the most massive cores and for cores with evidence of outflows. The PPMAP method is expected to result in a more accurate estimate for the mass-averaged dust temperatures than a straightforward colour temperature. However, our color temperature is calculated on the precise footprint of each core, with all the bands computed at the full 0.60\,$\arcsec$ resolution of our synthetic ALMA maps, while the PPMAP method yields a rather smooth temperature map at a nominal resolution of 2.5\,$\arcsec$. The distributions of $T_{\rm c}$ for all the synthetic ALMA cores are plotted in Figure~\ref{dust_models_histogram}, where the shaded blue histogram is for the WD model, and the shaded red histogram for the OH model. The color temperatures are clearly shifted to larger values than the dust temperatures of the candidate 3D cores, with median values of 20.4\,K and 16.0\,K for the WD and OH models respectively. The color temperatures are higher because the prestellar cores are the coldest regions along the line of sight and because the color temperature is most sensitive to the warmest regions that emit more strongly per unit mass.

Because the color temperature can be estimated observationally, it is useful to derive an empirical relation between $T_{\rm dust}$ and $T_{\rm c}$. Figure~\ref{dust_models_scatter} shows a scatter plot of $T_{\rm dust}$ versus $T_{\rm c}$ for our simulated cores (candidate 3D cores and corresponding {\sl getsf} cores, respectively), based on the WD dust model (empty blue circles) and on the OH model (filled red circles). The vertical scatter is quite large, particularly in the case of the OH model, but it is possible to fit linear relations in the log-log space, shown by the solid and dash-dotted lines in Figure~\ref{dust_models_scatter}. For the WD model the relation is almost exactly linear, with the dust temperature typically 70\% of the color temperature, $T_{\rm dust,WD}=7.2\,{\rm K}\times(T_{\rm c,WD}/10\,{\rm K})^{0.91}$. For the OH model the fit is significantly shallower than linear, with the dust temperature typically 50\% of the color temperature, $T_{\rm dust,OH}=5.6\,{\rm K}\times(T_{\rm c,OH}/10\,{\rm K})^{0.72}$. These empirical relations hold for color temperature maps at comparable angular resolution as the ALMA maps. This is not the case when combining for example the ALMA maps with Herschel images. In that case, since the color temperature at larger scales has a greater contribution from the warmer background, the typical ratio between $T_{\rm c}$ and $T_{\rm dust}$ will be larger than reported here. Figure~\ref{dust_models_scatter_Herschel} shows the case where $T_{\rm c}$ is computed from images at an angular resolution of 18$\arcsec$, corresponding to the 250\,$\mu$m band of Herschel. As expected, the total range of $T_{\rm c}$ is decreased compared to the higher-resolution case in Figure~\ref{dust_models_scatter}. However, the mean relation between $T_{\rm c}$ and $T_{\rm dust}$ expressed by the linear fit barely changes in the case of the WD dust model, 
\begin{equation}
T_{\rm dust,WD}=6.8\,{\rm K}\times(T_{\rm c,WD}/10\,{\rm K})^{0.91}.
\label{t_dust_col_wd}
\end{equation}
The change is instead significant for the OH model,
\begin{equation}
T_{\rm dust,OH}=5.5\,{\rm K}\times(T_{\rm c,OH}/10\,{\rm K})^{0.65}, 
\label{t_dust_col_oh}
\end{equation}
due to the larger temperature variations as a function of density caused by the higher opacity relative to the WD model, which is further discussed below. Equations~\ref{t_dust_col_wd} and \ref{t_dust_col_oh} may be used to convert estimated color temperature maps from Herschel observations into dust temperatures of cores extracted from ALMA observations, though the conversion should be derived again for different ALMA angular resolution or different distance of the observed region.

  \begin{figure}
   \centering
   \includegraphics[width=\hsize]{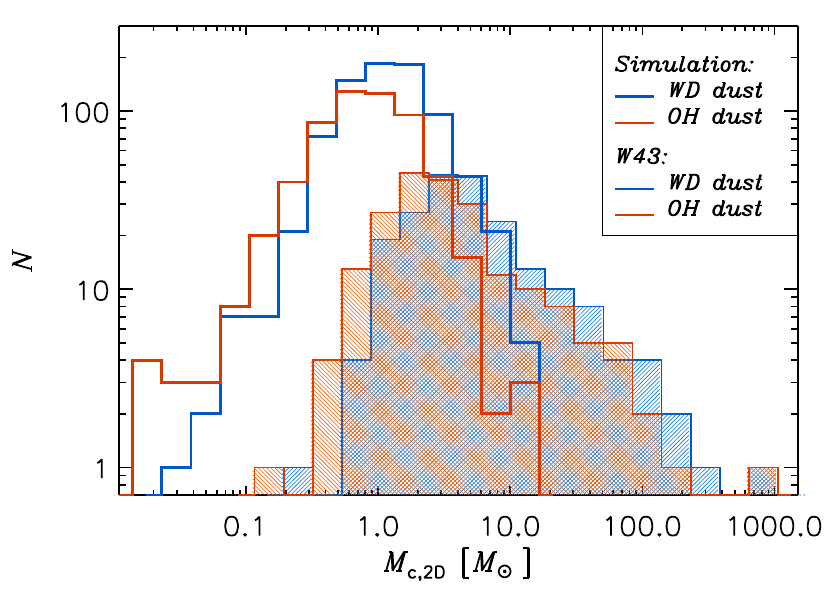}
      \caption{CMFs of the synthetic ALMA cores (unshaded histograms) from radiative transfer calculations with the WD dust model (blue histogram) and the OH dust model (red histogram). The shaded histograms are the CMFs of the W43 cores with masses derived from the integrated 1.3\,mm fluxes in \citet{Pouteau+22} using the optically-thin approximation of Equation~\ref{mass} with $T_{\rm dust}=14.4$\,K and $\kappa_{\rm 1,3\,mm}=0.0037$\,g\,cm$^{-2}$, as in the case of the WD dust model (blue histogram), and with $T_{\rm dust}=9.8$\,K and $\kappa_{\rm 1,3\,mm}=0.0083$\,g\,cm$^{-2}$, as in the case of the OH dust model (red histogram). The histograms of the W43 cores have been shifted to the right by a factor of 1.1 for clarity.   
              }
         \label{dust_models}
   \end{figure}

Figure~\ref{dust_models} compares the CMFs of the synthetic ALMA cores extracted from the maps computed with the WD model (blue unshaded histogram) and with the OH model (red unshaded histogram). The core masses are derived from the integrated {\sl getsf} 1.3\,mm fluxes, using Equation~\ref{mass}, the $\kappa_{1.3\,{\rm mm}}$ value of each model, and the median dust temperature of the candidate 3D cores as explained in \S~\ref{getsf}, $T_{\rm dust}=13.0$\,K and 7.5\,K for the WD and OH models respectively. The CMF based on the OH model is clearly shifted towards lower masses, by nearly a factor of two, relative to the CMF from the WD model. This shift can be understood as the effect of the higher opacity of the OH model, resulting in higher temperatures in the diffuse medium and lower in dense cores in comparison to the WD model. The stronger dependence of the dust temperature on density in the OH model causes a stronger background subtraction, hence a lower core mass estimate, than in the WD model. This is related to the rim-brightening effects discussed in \citet[][see Section 5.2 and Appendix B]{Menshchikov16}. Other sources of uncertainties in the conversion of integrated fluxes to masses, due to spatial variations in temperature and dust-opacity, have been discussed in the literature \citep[e.g.][]{Malinen+11,Roy+14,Pagani+15,Menshchikov16}. 

The radiative transfer calculations result in self-consistent dust temperature and opacity values, depending on the adopted dust model. Given that both are quite difficult to establish observationally, one may interpret the observations based on the temperature and opacity values from the simulation. This is shown in Figure~\ref{dust_models}, where the shaded histograms are the CMFs of the MM2 and MM3 regions of W43 derived from the integrated 1.3\,mm fluxes in \citet{Pouteau+22}, and the core masses are computed with the same pairs of values of $\kappa_{1.3\,{\rm mm}}$ and $T_{\rm dust}$ of the two dust models as in the case of the synthetic CMFs. The blue shaded histogram is based on the WD model, and the red shaded histogram on the OH model. The OH model causes a similar shift to lower masses relative to the WD model as in the case of the synthetic CMFs, but in this case the shift has a trivial origin, because the two CMFs are based on the same core extraction (the same integrated 1.3\,mm fluxes): the ratio of the two opacities is larger than the ratio of the two median temperatures, so the larger-opacity OH model results in lower masses. 

Besides the trivial dependence of the W43 CMF on the dust models, the more important result is the significant shift of the CMF to larger masses relative to the synthetic CMF and to the CMF in \citet{Pouteau+22}, where typical dust temperatures in excess of 20\,K are derived, while assuming a high dust opacity, $\kappa_{1.3\,{\rm mm}}=0.01\,{\rm cm^2\,g^{-1}}$. Based on our simulation and radiative-transfer calculations, and assuming that the cores are mostly prestellar as in our case, hence externally heated, such opacity value should yield rather low temperatures and hence larger masses than in \citet{Pouteau+22}. Temperatures below 10\,K and even down to 6\,K, have been observed in low-mass prestellar cores, based on spectral line observations of regions where the gas and dust temperatures should be closely coupled \citep{Crapsi2007,Harju2008}. Similarly low temperatures have been found in radiative transfer models, where (depending on the dust properties) the dust temperatures fall below 10\,K already in regions that are shielded by less than $A_{\rm V}=10$\,mag of extinction \citep{Zucconi2001, Stamatellos2007,ChaconTanarro2019}. Alternatively, if most of the W43 cores were protostellar, internal heating from the protostars may cause higher dust temperatures closer to those estimated in \citet{Pouteau+22}, but then the implications of the derived CMF for the origin of the stellar IMF would no longer be relevant.

\section{Conclusions} \label{conclusions}

We have identified and studied 12 regions forming high-mass stars in our 250\,pc star-formation simulation, driven self-consistently by SNe for over 40\,Myr. We have generated synthetic ALMA 1.3\,mm maps of these regions, by solving the radiative transfer with thousands of point sources, including all stars with mass $>2\,M_{\odot}$ formed self-consistently in the simulation. The synthetic ALMA images have been processed by following the same analysis pipeline and source-extraction code as in the ALMA-IMF Large Program. The cores extracted from the synthetic images with the {\sl getsf} code have been compared with a sample of cores extracted independently with the {\sl Dendro} code from the 2\,pc\,$\times$\,2\,pc\,$\times$\,250\,pc volumes used to generate the maps. The comparison shows that the masses of the cores from the synthetic maps are only very weakly correlated to those of the corresponding 3D cores, while the shapes and slopes of the CMFs remain roughly similar, despite the fact that the CMF from the synthetic maps is incomplete at all masses relative to the CMF from the corresponding 3D volumes.   

By first analyzing column-density maps, we have studied the role of projection effects while excluding other effects from radiative transfer, interferometer artifacts and thermal noise. The analysis shows that a significant fraction of the 2D {\sl getsf} cores are projection artifacts, an even larger fraction of 3D {\sl Dendro} cores are not retrieved in 2D, and the ones that are retrieved in 2D have estimated masses with large uncertainties. The following is a list of our main conclusions with respect to the core detection from column-density maps at a resolution of $\sim0.01$~pc, comparable to that of the ALMA-IMF project:

\begin{enumerate}

    \item The mass of the 2D {\sl getsf} cores is strongly correlated with the mass of the main density structure in their line of sight, showing that the background subtraction of {\sl getsf} works well in column-density maps despite the complex density field, although the extracted cores may not be real cores in 3D. 
    
    \item The fraction of projection artifacts, 2D {\sl getsf} cores without a 3D {\sl Dendro} counterpart, is a nearly monotonic function of mass, growing with decreasing mass from approximately 5\% above 2\,$M_{\odot}$, to over 50\% below 0.1\,$M_{\odot}$.
    
    \item The completeness fraction, 3D {\sl Dendro} cores with a 2D {\sl getsf} counterpart, increases monotonically with increasing mass without reaching unity. Above 1\,$M_{\odot}$, approximately 40\% of the 3D-selected cores are recovered in 2D.  
 
    \item The mass of the 2D {\sl getsf} cores is only a rough estimate of that of the associated 3D cores, with an uncertainty of approximately a factor of four, and a tendency to increasingly underestimate the 3D core mass towards increasing masses.
    
    \item The slope of the 2D CMF above 1\,$M_{\odot}$ is $-1.6\pm0.1$, steeper than that of the 3D CMF, $-1.2\pm0.1$.
      
\end{enumerate}   

When including effects from radiative transfer, interferometric artifacts, and thermal noise by using the synthetic 1.3\,mm ALMA maps, the ability to retrieve the correct core masses worsens significantly compared to the idealized case of the column-density maps. Our main conclusions with respect to the selection of cores in the synthetic ALMA maps, with comparable resolution and the same image processing and analysis as in the ALMA-IMF Large Program, are listed in the following:  

\begin{enumerate}
\setcounter{enumi}{5}
    
    \item Overall, 70\% of the synthetic ALMA cores have a 3D counterpart. Above 3\,$M_{\odot}$ only 10\% of them can be considered projection artifacts.
    
    \item A rather small fraction of the 3D {\sl Dendro} cores are detected as {\sl getsf} cores in the synthetic ALMA maps. The completeness fraction decreases monotonically with decreasing mass at all masses, from 40\% at 10\,$M_{\odot}$ to 15\% at 1\,$M_{\odot}$.

    \item When a 3D counterpart is found, {\em there is only a weak correlation between the masses of the synthetic ALMA cores and those of the corresponding 3D cores.} Interferometric artifacts, radiative transfer effects, and thermal noise cause a significant increase in the random error of the estimated core masses, relative to the idealized case of the column-density maps.   
    
    \item The slope of the CMF of the synthetic ALMA cores above 1~$M_{\odot}$, $-1.4\pm0.1$, is shallower than that of the cores selected from the column-density maps at the same spatial resolution, $-1.6\pm0.1$ (while still steeper than that of the 3D CMF, $-1.2\pm0.1$). This is primarily due to the combined effects of radiative transfer (spatial variations of the dust temperature) and interferometric artifacts (incomplete sampling of the (u,v) plane).

    \item Interferometric artifacts are found to cause a systematic shift in the derived masses of the synthetic ALMA cores, relative to those of the 2D {\sl getsf} cores from the column-density maps. A systematic shift of a similar magnitude is also caused by the choice of different dust models.
    
    \item The color temperature overestimates the real dust temperature in the cores. Empirical relations are derived to convert color temperatures from Hershel observations into dust temperature of cores extracted from ALMA observations.  
    
\end{enumerate}

Guided by the results of this analysis, we conclude that core masses and CMFs from current ALMA observations of Galactic protoclusters at kpc distances should not be taken at face value. With the combined use of observations and realistic simulations, it should eventually be possible to set strong constraints on the CMF, but it seems premature at this stage to draw conclusions on the origin of the stellar IMF directly from the observational results. We have shown that the observed CMF is expected to be a bit steeper than that of the real 3D cores, for the specific observational strategy, data-analysis pipeline, and source distances of the ALMA-IMF Large Program, and for star-forming regions like those found in our simulation. A general calibration of the relation between the real and observed CMFs will require a systematic study of the effects of spatial resolution and $(u,v)$ plane sampling and interpolation methods, as well as a set of simulated star-forming regions covering a wide range of physical conditions.

\section*{Acknowledgements}
Use of {\em getsf}, developed by Alexander Men’shchikov at the DAp, IRFU, CEA Saclay, France, is hereby acknowledged. We also thank Alexander Men’shchikov for his guidance in the use of {\em getsf}. We are grateful to Timea Csengeri for providing an updated version of the Table~4 in \citet{Csengeri+2017} used to produce our Figure~\ref{mass_size}, and to Yohan Pouteau and Fr\'{e}d\'{e}rique Motte for clarifications of some aspects of the ALMA-IMF Large Program data-analysis pipeline. Several comments by the referee and by members of the ALMA-IMF Large Program have helped us improve the paper.   
PP and VMP acknowledge financial support by the grant PID2020-115892GB-I00, funded by MCIN/AEI/ 10.13039/501100011033 and by the grant CEX2019-000918-M funded by MCIN/AEI/ 10.13039/501100011033. 
MJ acknowledges support from the Academy of Finland grant No. 348342.
The research leading to these results has received funding from the Independent Research Fund Denmark through grant No. DFF 8021-00350B (TH).
We acknowledge PRACE for awarding us access to Joliot-Curie at GENCI@CEA, France.
The astrophysics HPC facility at the University of Copenhagen, supported by research grants from the Carlsberg, Novo, and Villum foundations, was used for carrying out the postprocessing, analysis, and long-term storage of the results.

\section*{Data Availability}

Supplemental material can be obtained from a dedicated public URL (http://www.erda.dk/vgrid/alma-imf/).

\bibliographystyle{mnras}
\bibliography{MC,padoan,larson_relation}

\end{document}